\documentclass{elsarticle}
\makeatletter
\def\ps@pprintTitle{%
	\let\@oddhead\@empty
	\let\@evenhead\@empty
	\let\@oddfoot\@empty
	\let\@evenfoot\@oddfoot
}
\makeatother
\usepackage[margin=1in]{geometry}

\usepackage{bbm}
\usepackage{graphicx}
\usepackage{pstool}
\usepackage{wrapfig}
\usepackage{caption}
\usepackage{subcaption}
\usepackage[section]{placeins} 

\usepackage{fancyhdr} 
\usepackage{lastpage} 
\usepackage{extramarks} 

\usepackage{xcolor} 
\usepackage{enumerate} 
\usepackage{paralist} 
\usepackage{amsmath, amsthm, amssymb, mathtools}
\usepackage{mathabx, pifont, stmaryrd} 
\usepackage[explicit]{titlesec} 
\usepackage{etoolbox} 
\usepackage{bibentry} 
\makeatletter\let\saved@bibitem\@bibitem\makeatother 
\usepackage[colorlinks, bookmarksopen, bookmarksnumbered,
            citecolor=red,urlcolor=red]{hyperref} 
\makeatletter\let\@bibitem\saved@bibitem\makeatother 
\usepackage{todonotes}

\usepackage{algorithm}
\usepackage{algorithmic}

\newtheorem{remark}{Remark}

\theoremstyle{definition}

\usepackage{bm} 
\usepackage{amsfonts} 

\newcommand{\optuncm}[2]{\underset{#1}{\text{maximize}} ~~ #2}

\newcommand{\argoptunc}[2]{\underset{#1}{\arg\min} ~~ #2}

\newcommand{\pder}[2]{\ensuremath{\frac{\partial #1}{\partial #2}}}

\newcommand{\Acal}{\ensuremath{\mathcal{A}}}

\newcommand{\Dcal}{\ensuremath{\mathcal{D}}}

\newcommand{\Fcal}{\ensuremath{\mathcal{F}}}

\newcommand{\Ncal}{\ensuremath{\mathcal{N}}}

\newcommand{\Scal}{\ensuremath{\mathcal{S}}}

\newcommand{\Ucal}{\ensuremath{\mathcal{U}}}
\newcommand{\Vcal}{\ensuremath{\mathcal{V}}}
\newcommand{\Wcal}{\ensuremath{\mathcal{W}}}

\newcommand{\Ebb}{\ensuremath{\mathbb{E}}}

\newcommand{\Rbb}{\ensuremath{\mathbb{R} }}

\newcommand\Ibm{{\ensuremath{\bm{I}}}}

\newcommand\abm{{\ensuremath{\bm{a}}}}
\newcommand\bbm{{\ensuremath{\bm{b}}}}

\newcommand\nbm{{\ensuremath{\bm{n}}}}

\newcommand\sbm{{\ensuremath{\bm{s}}}}

\newcommand\xbold{\ensuremath{\mathbf{x}}}

\newcommand\mubold{{\ensuremath{\boldsymbol{\mu}}}}

\newcommand\phibold{{\ensuremath{\boldsymbol{\phi}}}}

\newcommand\psibold{{\ensuremath{\boldsymbol{\psi}}}}

\newcommand\epsilonbold{{\ensuremath{\boldsymbol{\epsilon}}}}
\newcommand\thetabold{{\ensuremath{\boldsymbol{\theta}}}}

\newcommand\Phibold{{\ensuremath{\boldsymbol{\Phi}}}}

\newcommand\Psibold{{\ensuremath{\boldsymbol{\Psi}}}}

\usepackage{tikz}
\usepackage{pgfplots}
\usepackage{pgfplotstable, filecontents, booktabs}
\pgfplotsset{compat=1.9}

\usetikzlibrary{pgfplots.groupplots}
\usepgfplotslibrary{fillbetween}
\usetikzlibrary{calc,fit,matrix,arrows,automata,positioning,shapes}
\usetikzlibrary{arrows.meta}

\pgfplotsset{select coords between index/.style 2 args={
    x filter/.code={
        \ifnum\coordindex<#1\fi
        \ifnum\coordindex>#2\fi
    }
}}

\tikzset{
 invisible/.style={opacity=0},
 visible on/.style={alt={#1{}{invisible}}},
 alt/.code args={<#1>#2#3}{%
   \alt<#1>{\pgfkeysalso{#2}}{\pgfkeysalso{#3}}
 },
}

\usepackage{setspace}
\usepackage{enumitem}
\usepackage{ulem}
\usepackage{makecell}
\usepackage{color}
\usepackage{tikz}
\usetikzlibrary{shapes}

\newbool{fastcompile}
\setbool{fastcompile}{true}

\newcommand\blfootnote[1]{%
	\begingroup
	\renewcommand\thefootnote{}\footnote{#1}%
	\addtocounter{footnote}{-1}%
	\endgroup
}

\begin{document}
\title{Bayesian conditional diffusion models for versatile spatiotemporal turbulence generation}

\author[ndAME,seas]{Han Gao\fnref{1st}}
\author[tuftsCS]{Xu Han\fnref{1st}}
\author[ndAME]{Xiantao Fan}
\author[ndAME,livemore]{Luning Sun}
\author[tuftsCS]{Li-Ping Liu}
\author[OSUMAE]{Lian Duan}
\author[ndAME,Lucy]{Jian-Xun Wang\corref{cor}}

\address[ndAME]{Aerospace and Mechanical Engineering Department, University of Notre Dame, Notre Dame, IN, USA}
\address[Lucy]{Lucy Family Institute for Data \& Society, University of Notre Dame, Notre Dame, IN, USA}
\address[seas]{School of Engineering and Applied Science, Harvard University, Cambridge, MA, USA}
\address[tuftsCS]{Department of Computer Science, Tufts University, Medford, MA, USA}
\address[livemore]{Lawrence Livermore National Laboratory, Livermore, CA, USA}
\address[OSUMAE]{Mechanical and Aerospace Engineering Department, The Ohio State University, Columbus, OH, USA}
\fntext[1st]{H.G(hgao1@seas.harvard.edu) and X.H(Xu.Han@tufts.edu) contributed equally.}
\cortext[cor]{Corresponding author: Jian-Xun Wang (jwang33@nd.edu)}


\begin{abstract}
Turbulent flows, characterized by their chaotic and stochastic nature, have historically presented formidable challenges to predictive computational modeling. Traditional eddy-resolved numerical simulations often require vast computational resources, making them impractical or infeasible for numerous engineering applications. As an alternative, deep learning-based surrogate models have emerged, offering data-drive solutions. However, these are typically constructed within deterministic settings, leading to shortfall in capturing the innate chaotic and stochastic behaviors of turbulent dynamics. In this study, we introduce a novel generative framework grounded in probabilistic diffusion models for versatile generation of spatiotemporal turbulence under various conditions. Our method unifies both unconditional and conditional sampling strategies within a Bayesian framework, which can accommodate diverse conditioning scenarios, including those with a direct differentiable link between specified conditions and generated unsteady flow outcomes, as well as scenarios lacking such explicit correlations. A notable feature of our approach is the method proposed for long-span flow sequence generation, which is based on autoregressive gradient-based conditional sampling, eliminating the need for cumbersome retraining processes. We evaluate and showcase the versatile turbulence generation capability of our framework through a suite of numerical experiments, including: 1) the synthesis of Large Eddy Simulations (LES) simulated instantaneous flow sequences from unsteady Reynolds-Averaged Navier-Stokes (URANS) inputs; 2) holistic generation of inhomogeneous, anisotropic wall-bounded turbulence, whether from given initial conditions, prescribed turbulence statistics, or entirely from scratch; 3) super-resolved generation of high-speed turbulent boundary layer flows from low-resolution data across a range of input resolutions. Collectively, our numerical experiments highlight the merit and transformative potential of the proposed methods, making a significant advance in the field of turbulence generation. \blfootnote{Videos of all numerical experiments can be found at \href{https://sites.nd.edu/jianxun-wang/bayesian-conditional-diffusion-models-for-versatile-spatiotemporal-turbulence-generation/}{https://sites.nd.edu/jianxun-wang/animations}} 

\end{abstract}
    
\maketitle
\section{Introduction}
\label{sec:intro}

Turbulent flows, ubiquitous in diverse contexts such as high-speed aircraft operation, oceanic currents, and combustion processes, exhibit complex unsteady and chaotic behaviors, characterized by swirling vortices and eddies over a wide spectrum of scales. To investigate and simulate these intricate phenomena, researchers often resort to numerical solutions of the governing partial differential equations (PDEs) of fluid dynamics. Eddy-resolved numerical simulations, notably Direct Numerical Simulation (DNS) and Large Eddy Simulation (LES), aims to accurately capture the unsteady/chaotic dynamics inherent in turbulent structures across extensive spatiotemporal scales. However, the computational demands of such methods can easily escalate to levels that are prohibitively expensive or practically infeasible. This presents significant challenges, particularly for tasks requiring rapid turnarounds, like real-time forecasting, or those that necessitating repeated model evaluations, as observed in design optimization and uncertainty quantification. 

Recent advances in deep learning (DL) offer new perspectives in addressing the intrinsic challenges of simulating unsteady turbulent flows. Over the past few years, there has been a surge of interest in harnessing the power of DL to enhance computational fluid dynamics (CFD) capabilities~\cite{brunton2020machine,vinuesa2022enhancing}. In this context, various DL-augmented CFD frameworks have emerged, wherein deep neural networks (DNNs) are integrated with conventional CFD algorithms, enabling functionalities such as discovering high-order discretizations~\cite{bar2019learning,kochkov2021machine}, refining coarse-grid computations~\cite{liu2022predicting,fan2023differentiable}, learning turbulence closure models~\cite{wang2017physics,maulik2019subgrid,maulik2021turbulent}, and accelerating pressure projection solvers~\cite{tompson2017accelerating}. Beyond enhancing existing CFD techniques, DL plays a pivotal role in creating rapid surrogate or reduced-order models, which serve as substitutes for the computationally-intensive numerical solvers for emulating spatiotemporal flow physics~\cite{raissi2019deep}. These innovations underscore the potential of DL in enhancing the accuracy and computational efficiency of fluid simulations. Despite these advancements, several challenges persist when it comes to turbulence simulation and prediction. A major concern is the reliability and robustness of DL-based solution, especially over a long time span. The chaotic nature of turbulence means that even minor perturbations or modeling inaccuracies can lead to significant deviations in long-term predictions. Thus, the deterministic nature of many DL-based fluid models often falls short in addressing the stochasticity intrinsic to turbulent flows. 

In response, generative modeling, grounded in probabilistic frameworks, emerges as a promising alternative. In the realm of data science, generative models encompass a set of algorithms designed to distill complex distributions of data. By leveraging latent representations~\cite{blei2003latent}, adversarial learning schemes~\cite{NIPS2014_5ca3e9b1}, or sequential data generation techniques~ \cite{kobyzev2020normalizing}, these models provide a comprehensive toolkit for a variety of academic and industrial applications, including but not limited to image generation, language processing, and anomaly detection~\cite{theis2015note}. The core of generative models lies in the objective of learning the underlying probability distributions of the data, which allows the synthesis of new data samples that adhere to the statistical properties inherent in the training set. In the context of turbulent flows, this implies the ability to generate instantaneous flow field realizations that mirror the same statistical characteristics of the observed turbulence data, thereby bypassing the need for costly CFD simulations. In recent years, there has been a growing interest in developing DL-based generative models for turbulent flows, motivated by their potential in flow reconstruction and super-resolution, synthetic inflow generation, and surrogate turbulent flow emulation. These existing methods can be broadly classified into following categories,

\underline{Autoregressive sequence models.} 
Existing DL-based methods for turbulence generation are mainly based on autoregressive learning architecture. These methods aim to learn the dynamic evolution of turbulent flow via neural networks based on labeled data. After training, these models can take flow fields from preceding time step as input, outputting flow predictions for subsequent steps. By rolling out the trained model given specified initial conditions, the sequence of turbulent flow fields can be synthesized in an autoregressive manner. These approaches mainly rely on the learned temporal correlations within the flow data, enabling the DL models to operate similarly to conventional explicit numerical solver. To address the challenge of handing high-dimensional turbulent flow data, these methods often incorporate dimensionality reduction techniques, such as Proper Orthogonal Decomposition (POD), Convolutional neural network (CNN) autoencoders, in conjunction with sequence networks. For instance, Fukami et al. ~\cite{fukami2019synthetic,fukami2019super,fukami2021machine} developed convolutional autoencoder-based autoregressive learning models for inflow turbulence synthesis, super-resolution or surrogate flow predictions. Yousif et al.~\cite{yousif2022physics} combined CNN autoencoder with LSTM to achieve the similar goals, and their models have been further extended by introducing adversarial training and attention mechanisms~\cite{yousif2023transformer}. However, it's important to emphasize that these autoregressive neural forecasting models intrinsically operate deterministically as they do not learn the underlying distribution of turbulence data, rendering them incapable of arbitrarily generating instantaneous turbulent flow realizations as a stochastic process. Furthermore, such deterministic autoregressive models are susceptible to cumulative error propagation, compromising the robustness of their long-term rollouts, with predictions either escalating unpredictably or dampening important flow features converged to time-averaged values. The inherently chaotic nature of turbulence exacerbates this vulnerability, leading to failures for long-term predictions. Uncertainty propagation in these models have been recently explored~\cite{maulik2020probabilistic,morimoto2022assessments}. 

\underline{GAN-based generative models.} 
To actually learn the underlying probability distribution of turbulent flow data, allowing for the random sampling of new realizations, Generative Adversarial Networks (GANs) have emerged as an effective tool. GANs rely on a competitive dynamic between generator and discriminator: the former generates synthetic turbulent data, while the latter differentiates synthetic from labeled data. Through iterative adversarial training, the outputs of the generator are progressively refined, targeting convergence to the true distribution of the training data. Recent studies have demonstrated the applicability of GANs in turbulent flow generation, super-resolution and inpainting. Several GAN variants, including the Wasserstein GAN (WGAN), conditional GAN (cGAN), and deep convolutional GAN (DCGAN), have been adapted to synthesize individual snapshots of both homogeneous isotropic turbulence and wall-bounded turbulent flows~\cite{stengel2020adversarial,drygala2022generative}. In the super-resolution (SR) context, methods such as super-resolution GAN (SRGAN), enhanced SRGAN, and cycle-consistent GAN (CycGAN) have been employed to enhance the low-resolution or low-fidelity 2D/3D turbulence snapshots~\cite{deng2019super,kim2021unsupervised,guemes2021coarse,yu2022three}. Buzzicotti et al.~\cite{buzzicotti2021reconstruction} leverage GAN to generate missing turbulence data similar to image impainting. 
However, these GAN-based models primarily focus on single-snapshot generation Since such GANs are trained on isolated flow snapshots (treated as 2D or 3D spatial images without temporal coherence), they intrinsically lack the capacity for sequential saptiotemporal turbulent flow synthesis. While certain research endeavors have integrated GAN with sequential networks~\cite{yousif2023transformer}; however, these GANs largely serve as deterministic encoders, with their adversarial training paradigm remaining decoupled from sequential networks, limiting their true stochastic generative potential. Only a few studies have rigorously explored the potential of GANs to approximate the probabilistic distribution of flow sequence data without input-output labels for training. Notably, Xie et al.\cite{xie2018tempogan} proposed TempoGAN, a combination of GAN and RNN, designed to generate stochastic, temporally consistent fluid simulations with SR capability. Similar to TempoGAN, Kim and Lee~\cite{kim2020deep} developed a combined WGAN+RNN model to generate 2D inlet turbulence for 3D channel flow simulations. Despite the promise of GANs in turbulent flow synthesis, they face several challenges: (1) the training of GAN can often be notoriously unstable, leading to frequent oscillations between the generator and discriminator~\cite{gui2021review}; (2) moreover, GANs also often suffer from ``mode collapse'', producing limited varieties of outputs~\cite{bau2019seeing}. 

\underline{Normalizing flow (NF)-based generative models.} 
Normalizing flows (NFs) have emerged as another notable subclass of generative models. At their core, NFs transform a basic, predefined probability distribution through a series of invertible, differentiable functions, morphing it into a complex one that approximates the underlying data distribution. The strength of NFs is their potential to model a complex distribution directly, bypassing the need for sampling-based approximations. Conceptually, this process of transformation resembles a continuous flow, progressively shaping the initial distribution to align with the target data distribution. Geneva and Zabaras \cite{geneva2020multi} proposed a multi-fidelity deep generative model using NFs to simulate turbulent flows at different Reynolds numbers. Efficient low-fidelity solver is leveraged to generate high-fidelity solution, significantly reducing the computational costs. 
Their model incorporates a conditional invertible neural network with LSTM connections, trained through both data and governing equations using a variational loss function. Another recent study integrated an attention-based sequence model into the NF-based generative architecture, enabling the probabilistic synthesis of time-evolving turbulence~\cite{sun2023unifying}. Despite their promise, NFs are known to have scalability issues with high-dimensional data, like turbulent flows, due to the requirement of computing Jacobians for the transformations. Furthermore, developing conditional variants of NFs is notably challenging, limiting their adaptation to new turbulent flow conditions.

\underline{Diffuison-based generative models.} 
Diffusion models have recently marked a significant footprint in generative modeling, demonstrating superior capabilities compared to GANs or NFs in multiple computer vision applications~\cite{song2019generative,ho2020denoising,dhariwal2021diffusion}. These models typically involve a noise-addition process followed by a reverse denoising operation via deep neural networks. Categorically, diffusion models can be classified into (i) Denoising Diffusion Probabilistic Models (DDPMs)~\cite{ho2020denoising}, (ii) Score-based diffusion models~\cite{song2019generative}, and (iii) Stochastic Differential Equation (SDE) based models~\cite{song2020score}, with the latter considered as an overarching framework for the former two. Diffusion generative models offer three primary advantages. First, their training process is straightforward, demanding minimal architecture and hyperparameter fine-tuning. Second, their hidden variables mirror the physical characteristics of samples, enabling seamless multi-scale feature capturing. Lastly, operating within a Bayesian framework, they provide a more direct way to enable conditioned generation compared to normalizing flow-based models. Despite their evident success in fields like image generation and super-resolution, their application in the physics domain, particularly turbulence modeling, remains limited. Some recent works have made preliminary inroads on 2D Komogrov flows, targeting specific sub-tasks like super-resolution~\cite{shu2023physics,apte2023diffusion,wan2023debias}. However, a comprehensive application of diffusion models for spatiotemporal generation of inhomogeneous, anisotropic  turbulence, particularly via conditional frameworks, remains a largely uncharted territory.  A gap exists in systematically generating intricate turbulent flows under various conditional inputs, such as initial fields, boundary conditions, turbulent statistics, external forces, etc.

\begin{figure}[H]
	\centering
	\includegraphics[width=0.8\textwidth]{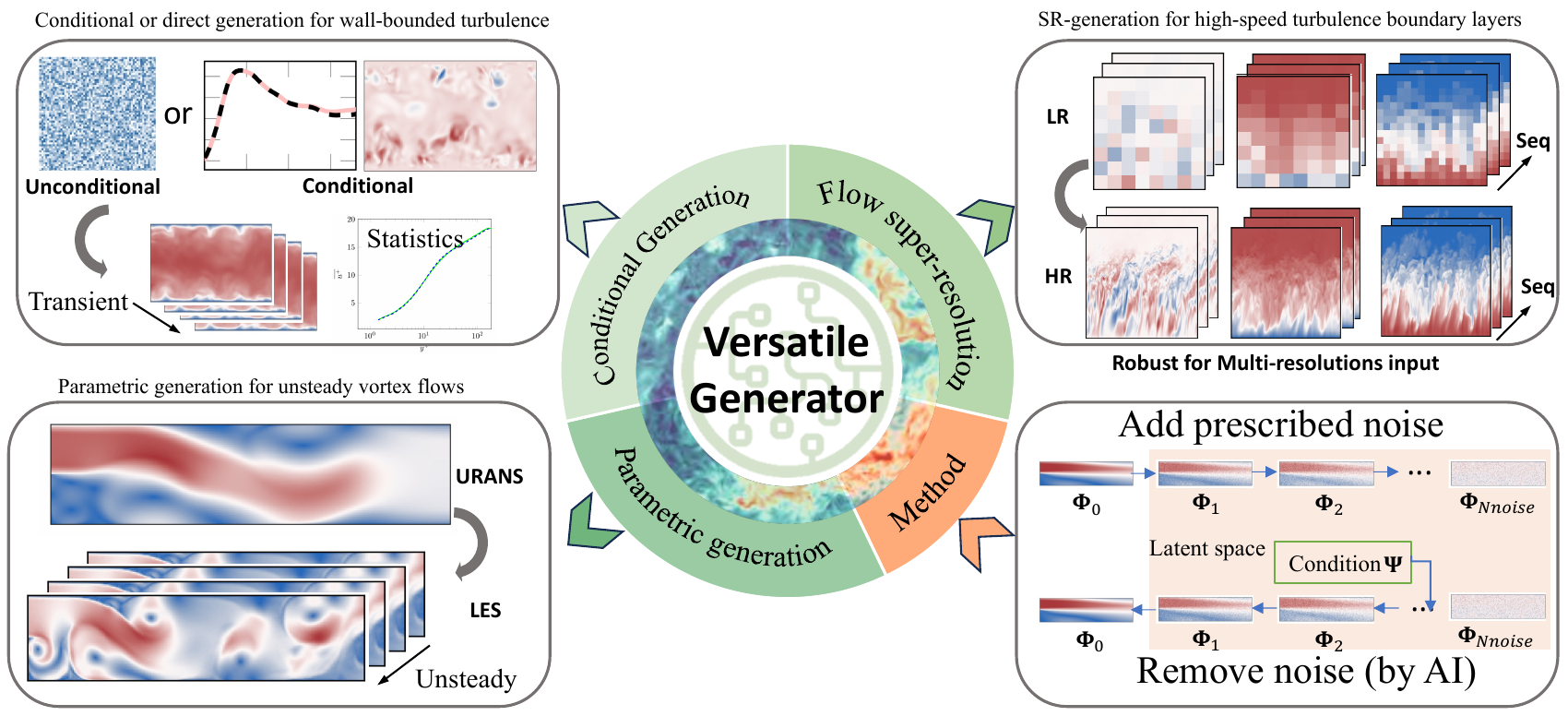}
	\caption{Overview of the proposed versatile spatiotemporal turbulence generator via conditional diffusion modeling.}
	\label{fig:overview}
\end{figure}

To this end, this work delves into the development and systematic investigation of conditional diffusion models designed for generating complex inhomogeneous and anisotropic turbulent flows under various conditions. Unlike the majority of existing literature on generative modeling that are primarily limited to single-snapshot generation, our emphasis lies in capturing the underlying probabilistic distribution of time-evolving turbulent flow sequences. Essentially, our diffusion model can be conceptualized as a stochastic spatiotemporal process, thereby allowing to generate new realizations of instantaneous turbulent flow sequences through randomly sampling. Furthermore, to enhance the versatility of our generator, we integrate advanced Bayesian conditional sampling techniques. By conditioning the generation (sampling) process on condition parameters such as initial flow fields, boundary conditions, turbulent statistics, and coarse-grained solutions, our model adeptly synthesizes turbulent flows tailored to specific constraints.  The proposed Bayesian conditional  sampling approach not only bridges parametric conditions with generated spatial-temporal turbulent flows, but also magnifies the model's flexibility, making it highly adaptable to a range of real-world scenarios. 
As depicted in Fig.~\ref{fig:overview}, the versatility of our proposed method is evident across a variety of turbulence generation scenarios. In the context of 2D unsteady vortex flows over a backward-facing step, the model can reproduce LES-like eddy-resolved turbulent given the corresponding low-fidelity URANS simulated solutions. For 3D turbulent channel flows, the figure emphasizes the model's capability at generating spatiotemporal sequences from varied initial conditions, specific flow statistics, or even entirely from scratch. Furthermore, the 3D compressible supersonic turbulent boundary layers scenario showcased in the figure showcase our model's ability in super-resolution generation—demonstrating a conditioned generation from low-resolution inputs to high-fidelity DNS of high-speed turbulent boundary layers. Collectively, Fig.~\ref{fig:overview} illustrate the robust adaptability and comprehensive capability of the proposed conditional diffusion model in addressing diverse turbulence generation challenges. 

The \emph{primary contributions} of the paper are summarized as follows: (a) We present a novel generative framework for spatiotemporal turbulent flow, rooted in probabilistic diffusion modeling. This model operates within the Bayesian context, trained using the evidence lower bound, and is designed to randomly generate temporarily-coherent turbulent flow snapshots over a long-time span. This approach sets it apart from other GAN-based methods that typically focus on generating single snapshots. (b) We proposed a systematic approach to conditional generation, addressing two distinct scenarios of conditioning. The first establishes a differentiable relationship between specified conditions and the corresponding spatial-temporal turbulence solutions. Notably, we leverage the gradient from this connection via automatic differentiation to directly modify the unconditional score functions, allowing conditiional sampling without necessitating model retraining. For the second scenario, where conditions lack a direct correlation to the turbulent flow, we treat them as additional inputs to our diffusion model. With transfer-retraining, the conditioned probability distribution can be learned from conditioned datasets. (c) We proposed autoregressive conditioned generation techniques that enables long-span flow generation.  Its efficacy is benchmarked against the replacement method by Ho et al.~\cite{ho2022video} designed for prolonged video creation. (d) Our framework, when conditioned on URANS simulations, is capable of generating LES-like solutions with unsteady dynamics that are not captured by URANS, including phenomena such as Kelvin–Helmholtz instability and vortex shedding. (e) Through a series of comprehensive experiments, we evaluate the performance of the proposed method in generating anisotropic, inhomogenous wall-bounded turbulence, under different given conditions. These generated flows, both from unconditional and gradient-based conditional sampling, are benchmarked against DNS results to affirm their fidelity and accuracy. (f) We also showcased our model's versatility through its application in super-resolution tasks for a high-speed turbulent boundary layer flow, with a particular emphasis on handling various input resolutions. 

The remainder of the paper is structured as follows. Section~\ref{sec:method} delves into the methodology of the our diffusion-model-based turbulence generator. Specifically, the formulation of the variational diffusion model for spatiotemporal turbulent flows is presented in Section~\ref{sec:VDF}, while the unconditional and conditional sampling methods are discussed in Sections~\ref{sec:unconditionalsample} and~\ref{sec:condisample}, respectively. The generation of extended flow sequences using autoregressive gradient-based conditional sampling, which circumvents re-training, is detailed in Section~\ref{sec:longvideo}. In Section~\ref{sec:results}, we present a comprehensive set of numerical experiments to demonstrate and evaluate the capability of the proposed method in diverse turbulence generation contexts. This section showcases: (1) the model's proficiency in simulating 2D unsteady flows over a backward-facing step, especially in generating LES-like turbulent flows from URANS solutions; (2) Its capability in generating wall-bounded turbulence, whether from specified initial conditions, statistical data, or from scratch; (3) Its ability in SR generation for compressible supersonic turbulent boundary layer flows, demonstrating resilience to diverse input resolutions. Finally, the paper is concluded in Section \ref{sec:conclusion}, summarizing our principal findings and projecting avenues for subsequent enhancements.

\section{Methodology}
\label{sec:method}

\subsection{Learning probability distribution of spatiotemporal turbulence via variational diffusion models}
\label{sec:VDF}
Turbulent flows are intrinsically characterized as stochastic spatiotemporal processes, reflecting their chaotic and random nature across spatial and temporal scales. In this context, a generative model aims to capture and reproduce these complex dynamics by learning the underlying probability distribution $p(\Phi_0)$ of the spatiotemporal turbulent flow state variables $\Phi_0(\xbold, t)$, which can be discretized into a sequence of spatial fields $\Phibold_0\in\Rbb^{N_\mathrm{length}\times N_\mathrm{dof}}$, encompassing $N_\mathrm{length}$ snapshots, with each having $N_\mathrm{dof}$ degree of freedom. Our objective, therefore, is to construct a learning model $p_\thetabold(\Phibold_0)$ that approximates the true probability density  $p(\Phibold_0)$ by learning from a training dataset $\Acal_\mathrm{train}\subset\Rbb^{N_\mathrm{length}\times N_\mathrm{dof}}$ that contains many realizations of $\Phibold_0$. This can be theoretically realized by maximizing the likelihood of all training realizations $\Phibold_0\in \Acal_\mathrm{train}$ as expressed by,
\begin{equation}
	\optuncm{\thetabold}{\sum_{\Phibold_0\in \Acal_\mathrm{train}} \log p_\thetabold(\Phibold_0)}, 
\end{equation}
where $\thetabold\in\Rbb^{N_\thetabold}$ represents the model parameter vector, such as those in neural networks, which is to be optimized. Specifically, we propose constructing a probabilistic diffusion model~\cite{yang2022diffusion} that implicitly learns this underlying distribution, enabling rapid sampling (generation) of new realizations of spatiotemporal turbulent flow fields.  To this end, we introduce a series of latent variables, $\Phibold_1,\dots \Phibold_{N_\mathrm{noise}}\in \Rbb^{N_\mathrm{length}\times N_\mathrm{dof}}$, which are generated from the turbulent flow sequence $\Phibold_0$ via a {Markovian} process, 
\begin{equation}
	q(\Phibold_{j}|\Phibold_{i}) \coloneqq \Ncal(\alpha_{j|i}\Phibold_{i},\sigma^2_{j|i}\Ibm),\quad \text{for any } 0 \leq i < j \leq N_\mathrm{noise}, 
	\label{eqn:forward_process_1step}
\end{equation}
where $\Ncal(\abm,\bbm)$ represents a multi-variant Gaussian distribution with the mean $\abm\in\Rbb^{N_\mathrm{length}\times N_\mathrm{dof}}$ and covariance matrix $\bbm\in\Rbb^{(N_\mathrm{length}\times N_\mathrm{dof})^2}$;  $\Ibm\in\Rbb^{(N_\mathrm{length}\times N_\mathrm{dof})^2}$ is the identity matrix; the set of scalar $\{\sigma_{j|i},\alpha_{j|i} \big| \text{for any } 0 \leq i < j \leq N_\mathrm{noise}\}$ are predefined such that the marginalized distribution  $p(\Phibold_{N_\mathrm{noise}}|\Phibold_0)$ converge to a zero-mean Gaussian distribution  $p(\Phibold_{N_\mathrm{noise}})=\Ncal(\mathbf{0},\sigma^2_{N_{\mathrm{noise}}|0}\Ibm)$, which is straightforward to sample~\cite{kingma2021variational}. The conditional distribution of each perturbed state given the turbulent flow sequence data $\Phibold_{0}$ is Gaussian,
\begin{equation}
	q(\Phibold_{i}|\Phibold_0) = \Ncal(\alpha_i\Phibold_0,\sigma_i^2\Ibm),
	\label{eqn:forward_process_istep}
\end{equation}
where $\{\alpha_i,\sigma_i \big| \text{for any } 1 \leq i \leq N_\mathrm{noise}\}$ are scalars that can be analytically derived from the the predefined set $\{\sigma_{j|i},\alpha_{j|i} \big| \text{for any } 0 \leq i < j \leq N_\mathrm{noise}\}$~\cite{luo2022understanding,ho2020denoising}. This process is known as the {\it forward diffusion process}, which is determined by handcrafting the transition kernels to progressively transform the data distribution $p(\Phibold_{0})$ into an accessible Gaussian distribution $p(\Phibold_{N_\mathrm{noise}})$. 

If this diffusion process can be inverted, such an {\it reverse diffusion process} would enable the synthesis of new realizations of turbulent flow sequences simply by sampling from the Gaussian distribution $p(\Phibold_{N_\mathrm{noise}})$. However, the reverse transition kernel $q(\Phibold_{i-1}|\Phibold_{i})$ is not known {\it a priori} and therefore necessitates learning through neural network parameterization. It is noteworthy that the reverse transition kernel becomes tractable when conditioned on $\Phibold_{0}$~\cite{song2020denoising,luo2022understanding},
\begin{equation}
	q(\Phibold_{i-1}|\Phibold_{i}) = q(\Phibold_{i-1}|\Phibold_{i},\Phibold_{0})=\frac{q(\Phibold_{i}|\Phibold_{i-1},\Phibold_{0})q(\Phibold_{i-1}|\Phibold_{0})}{q(\Phibold_{i}|\Phibold_{0})}\quad\text{for any } 2\leq i \leq N_\mathrm{noise},
	\label{eqn:transkernel}
\end{equation}
where the first equality is from the Markovian property and the second is based on the Bayes rule. If the perturbation step $\sigma_i^2$ is sufficiently small, the reverse transition kernel is also Gaussian, which can be analytically derived by substituting  $q(\Phibold_{i}|\Phibold_{i-1})$ from Eq.~\eqref{eqn:forward_process_1step}, and both $q(\Phibold_{i-1}|\Phibold_{0})$ and $q(\Phibold_{i}|\Phibold_{0})$ from \eqref{eqn:forward_process_istep}, 
\begin{equation}
		q(\Phibold_{i-1}|\Phibold_{i}) = q(\Phibold_{i-1}|\Phibold_{i},\Phibold_{0}) = \Ncal\Big( \Phibold_{i-1}; \tilde{\mubold}\big(\Phibold_i,\Phibold_0\big), \tilde{\sigma}_i^2\Ibm \Big), \quad\text{for any } 2\leq i \leq N_\mathrm{noise},
	\label{eqn:reverse_true}
\end{equation}
with mean $\tilde{\mubold}\big(\Phibold_i,\Phibold_0\big) = \frac{\alpha_{i|i-1}\sigma_{i-1}^2}{\sigma^2_i}\Phibold_i+\frac{\alpha_{i-1}\sigma^2_{i|i-1}}{\sigma_i^2}\Phibold_0$ and covariance $\tilde{\sigma}_i^2\Ibm = \frac{\sigma_{i|i-1}^2\sigma_{i-1}^2}{\sigma_i^2}\Ibm$~\cite{kingma2021variational}. Given that $\Phibold_0$ represent the generated flow outcomes, which are not known {\it a priori}, we approximate $q(\Phibold_{i-1}|\Phibold_{i},\Phibold_{0})$ by $p_\thetabold(\Phibold_{i-1}|\Phibold_{i})$ parameterized with trainable $\thetabold$, which can be optimized by maximizing the evidence lower bound (ELBO) given the training set $\Phibold_0\in \Acal_\mathrm{train}$.  The expression for ELBO is given as,
\begin{equation}
	\begin{split}
		\mathrm{ELBO}_\thetabold(\Phibold_{0}) = \ 
		& \Ebb_{q(\Phibold_1|\Phibold_0)}\Big[\log p_\thetabold(\Phibold_0|\Phibold_1)\Big] \\
		&-D_{\mathrm{KL}}\Big(
		q(\Phibold_{N_\mathrm{noise}}|\Phibold_0)
		\big|\big|
		{p}(\Phibold_{N_\mathrm{noise}})
		\Big)\\
		&-\sum_{i=2}^{N_\mathrm{noise}}\Ebb_{q(\Phibold_i|\Phibold_0)}	
		\Big[
		D_\mathrm{KL}\Big(
		q(\Phibold_{i-1}|\Phibold_{i},\Phibold_{0})
		\big|\big|
		p_\thetabold(\Phibold_{i-1}|\Phibold_{i})
		\Big)
		\Big].	
	\end{split}
	\label{eqn:ELBO}
\end{equation} 
Since the second term $-D_{\mathrm{KL}}\Big(q(\Phibold_{N_\mathrm{noise}}|\Phibold_0)\big|\big|{p}(\Phibold_{N_\mathrm{noise}})\Big)$ of Eq. \eqref{eqn:ELBO} is non-trainable and close to zero given Eq.~\eqref{eqn:forward_process_1step}, we only need to minimize the following terms,
\begin{equation}
		 - \Ebb_{q(\Phibold_1|\Phibold_0)}\Big[\log p_\thetabold(\Phibold_0|\Phibold_1)\Big] 
		+ \sum_{i=2}^{N_\mathrm{noise}}\Ebb_{q(\Phibold_i|\Phibold_0)}	
			\Big[
					D_\mathrm{KL}\Big(
					q(\Phibold_{i-1}|\Phibold_{i},\Phibold_{0})
					\big|\big|
					p_\thetabold(\Phibold_{i-1}|\Phibold_{i})
			\Big)
			\Big].
\label{eqn:ELBO}	
\end{equation}
To parameterize $p_\thetabold$, we use the Eq.~\ref{eqn:reverse_true}
\begin{equation}
	p_\thetabold(\Phibold_{i-1}|\Phibold_{i}) \coloneqq q\big(\Phibold_{i-1}|\Phibold_{i},\hat{\Phibold}_\thetabold(\Phibold_i;i)\big),
	\label{eqn:phihat1}
\end{equation}
where ${\Phibold}_{0}$ is parameterized using deep neural networks with trainable parameters $\thetabold$,
\begin{equation}
	{\Phibold}_0 \approx \hat{\Phibold}_\thetabold(\Phibold_i; i) = \Ebb_{\Phibold_0\sim p(\Phibold_0|\Phibold_i)}[\Phibold_0].
\label{eqn:phihati}
\end{equation}
In particular, from re-parametrization trick given by Eq.~\ref{eqn:forward_process_istep}, ${\Phibold}_{0}$ can be expressed as,
\begin{equation}
	\hat{\Phibold}_\thetabold(\Phibold_i; i) = \frac{\Phibold_{i} - \sigma_i \epsilonbold_{\thetabold}(\Phibold_i; i)}{\alpha_i}
\end{equation}
where the noise $ \epsilonbold_{\thetabold}(\Phibold_i; i)$ is parameterized by a U-Net variant with residual blocks, attention, and positional embedding as detailed in~\cite{song2020denoising,ho2022video}. Given that the KL divergence between two Gaussian distributions has a closed form, maximizing the ELBO defined in Eq.~\ref{eqn:ELBO} can be simplified to following optimization,
\begin{equation}
	\thetabold^* =	\argoptunc{\thetabold}{\sum_{\Phibold_{0}\in\Acal_\mathrm{train}}\Ebb_{\epsilonbold\sim \Ncal(\mathbf{0},\Ibm), i\sim\Ucal(1,N_\mathrm{noise})}\big[C_i\|\hat{\Phibold}_{\thetabold}(\Phibold_i;i)-\Phibold_0\|^2_{L2}\big]},
	\label{eqn:loss}
\end{equation}
where $C_i = \frac{\alpha_{i-1}^2\sigma^2_{i|i-1}}{2\sigma_{i-1}^2\sigma_{i}^2}$ for $i\geq2$ and $C_1=1$; $\Ucal(1,N_\mathrm{noise})$ is the discrete uniform distribution from 1 to $N_\mathrm{noise}$; $\Phibold_i$ is sampled using re-parametrization based on Eq.~\eqref{eqn:forward_process_istep}~ \cite{song2020denoising,ho2020denoising,kingma2021variational}. Namely, it can be computed by $\Phibold_i=\alpha_{i}{\Phibold}_{0}+\sigma_i\epsilonbold$, where $\epsilonbold$ is standard Gaussian noise, i.e., $\epsilonbold \sim \mathcal{N}(\bm{0}, I)$ and $\Phibold_0\in \Acal_\mathrm{train}$. 

\subsection{Unconditional generation of spatiotemporal turbulent flow sequences}
\label{sec:unconditionalsample}
After training the diffusion model, we can generate new spatiotemporal turbulent flow sequences. This is achieved by sampling the learned distribution via the reverse diffusion process, starting from a multivariate Gaussian distribution $\Phibold_{N_\mathrm{noise}} \sim \Ncal(\mathbf{0},\sigma^2_{N_{\mathrm{noise}}|0}\Ibm)$.  Specifically, we progressively sample $\Phibold_{i-1}$ based on $\Phibold_{i}$ using the learned reverse transition kernel,
\begin{equation}
	p_{\thetabold^*}(\Phibold_{i-1}|\Phibold_{i}) = \Ncal\Big(\tilde{\mubold}_{\thetabold^*}(\Phibold_i; i), \tilde{\sigma}_i^2 \Ibm\Big)\quad\text{for any } 2\leq i \leq N_\mathrm{noise},
	\label{eqn:ptheta_star}
\end{equation}
where  $\tilde{\sigma}_i^2 = \frac{\sigma_{i|i-1}^2\sigma_{i-1}^2}{\sigma_i^2}$ and $\tilde{\mubold}_{\thetabold^*}(\Phibold_i; i) = \frac{\alpha_{i|i-1}\sigma_{i-1}^2}{\sigma^2_i}\Phibold_i+\frac{\alpha_{i-1}\sigma^2_{i|i-1}}{\sigma_i^2}\hat{\Phibold}_{\thetabold^*}(\Phibold_i;i)$. The Eq.~\eqref{eqn:ptheta_star} illustrates how we can sample $\Phibold_{i-1}$ based on $\Phibold_i$ from a multivariate Gaussian with a mean of $\mubold_{\thetabold^*}(\Phibold_i; i)$. This process can be equivalently perceived as $\Phibold_i$ being denoised to move closer to $\hat{\Phibold}_{\thetabold^*}(\Phibold_i;i)$~\cite{kingma2021variational}. Namely, the mean of the learned reverse transition kernel can be expressed in terms of the neural network-based noise estimate $\epsilonbold_{\thetabold^*}(\Phibold_i;i)$ as,
\begin{equation}
	\tilde{\mubold}_{\thetabold^*}(\Phibold_i; i) = \frac{1}{\alpha_{i|i-1}}\Phibold_i-\frac{\sigma^2_{i|i-1}}{\alpha_{i|i-1}\sigma_i}\epsilonbold_{\thetabold^*}(\Phibold_i;i),
	\label{eqn:mutheta_star}
\end{equation}
To improve the quality of samples, a second-order noise correction is employed~\cite{ho2022video} as detailed,
\begin{equation}
	\epsilonbold^\mathrm{correct}_{\thetabold^*}(\Phibold_i;i) = \frac{1}{2}\Big(\epsilonbold_{\thetabold^*}(\Phibold_i; i)  + \epsilonbold_{\thetabold^*}(\tilde{\mubold}_{\thetabold^*}; i) \Big),
\end{equation}
Subsequently, by substituting the original noise estimate, $\epsilonbold_{\thetabold^*}(\Phibold_i;i)$ in Eq.~\eqref{eqn:mutheta_star} with $\epsilonbold^\mathrm{correct}_{\thetabold^*}(\Phibold_i;i)$, the corrected mean is derived as
\begin{equation}
	\tilde{\mubold}^\mathrm{correct}_{\thetabold^*}(\Phibold_i; i) = \frac{1}{\alpha_{i|i-1}}\Phibold_i-\frac{\sigma^2_{i|i-1}}{\alpha_{i|i-1}\sigma_i}\epsilonbold_{\thetabold^*}^\mathrm{correct}(\Phibold_i;i).
	\label{eqn:mutheta_correct_star}
\end{equation}
From Eq.~\eqref{eqn:ptheta_star} and Eq.~\eqref{eqn:mutheta_correct_star}, we introduce a mapping, $\Scal_{\mathrm{uc}}: (\Phibold_i,\epsilonbold)\mapsto\Phibold_{i-1}$
\begin{equation}
	\Phibold_{i-1} = \tilde{\mubold}_{\thetabold^*}^{\mathrm{correct}}(\Phibold_i; i) + \tilde{\sigma}_i\epsilonbold
	\label{eqn:Suc}
\end{equation}
where $\epsilonbold$ represent standard multivariate Gaussian noise. Iteratively applying Eq.~\eqref{eqn:Suc}, we can sample a new $\Phibold_0$ starting from $\Phibold_{N_\mathrm{noise}}$. To set the stage for the discussion on conditional generation in Section~\ref{sec:condisample}, we can reinterpret Eq.~\eqref{eqn:mutheta_star} through the lens of the score-based generative modeling framework~\cite{song2020score},
\begin{equation}
	\tilde{\mubold}_{\thetabold^*}(\Phibold_{i}; i) = \frac{1}{\alpha_{i|i-1}}\Phibold_i + \frac{\sigma^2_{i|i-1}}{\alpha_{i|i-1}}\sbm_{\thetabold^*}(\Phibold_{i}; i),
	\label{eqn:score_modeling}
\end{equation}
where
\begin{equation}
	\sbm_{\thetabold^*}(\Phibold_{i};i)  \coloneqq \nabla_{\Phibold_i}\log p(\Phibold_i) = \frac{\alpha_i\hat{\Phibold}_{\thetabold^*}(\Phibold_i;i) - \Phibold_i}{\sigma^2_i}
	\label{eqn:s_theta_star}
\end{equation}
is the score model that approximates the fastest increasing direction of the probability density within the state space, as known as stein score~\cite{song2019generative,song2020score}. The recursive unconditional sampling outlined in this section illustrates the progression from a point of low-probability-density, $\Phibold_{N_\mathrm{noise}}$, towards the region of high probability density, $\Phibold_{0}$.

\subsection{Generation of spatiotemporal turbulent flow sequences with conditions}
\label{sec:condisample}

The capability to generate turbulent flows becomes substantially more significant when tailored to specific parameters, prior knowledge or certain constraints. This conditional generation method enables the synthesis of turbulent flows for specified conditions or scenarios, augmenting its intrinsic value. For example, it will be very useful if we can use the data from efficient low-fidelity (LF) Unsteady Reynolds-Averaged Navier-Stokes (URANS) simulations as the condition to generate corresponding eddy-resolved instantaneous spatiotemporal turbulence solutions, which typically requires high-fidelity simulations, such as Large Eddy Simulations (LES) or Direct Numerical Simulations (DNS) that are computationally demanding. Moreover, when faced with low-resolution turbulence datasets -- either due to storage constraints or measurement limitations --  conditional sampling will  super-resolved generation of high-resolution data, recovering back high-frequency, high-wavenumber details of the flow. Another practical applications of the conditional generation is for data assimilation. Namely, it's advantageous for the generator to yield flow realizations in alignment with available measurement data, which are often sparse or indirect in nature. Building on the foundational methodology introduced in Section~\ref{sec:unconditionalsample}, this subsection delves into the conditional turbulence generation. 

Mathematically, any set of conditions can be parameterized as a vector, $\Psibold\in\Rbb^{N_\Psibold}$, referred as to the condition vector in this work. Examples of the condition vector $\Psibold$ include the URANS solution vector, instantaneous flow measurements, flow configuration parameters, low-resolution snapshots, among others. In the context of probabilistic generation modeling, taking into account these additional conditions in the generation process is mathematically equivalent to obtaining the conditional probability $p(\Phibold_0|\Psibold)$ of the state $\Phibold_0$ given the condition vector $\Psibold$. Similar to the unconditional generation discussed above, our objective here is to build a learning model $p_\thetabold(\Phibold_0|\Psibold)$ to approximate the true conditional probability distribution $p(\Phibold_0|\Psibold)$. From Bayes's rule, the learned conditional density can be expressed as,
\begin{equation}
	p_{\thetabold}(\Phibold_0|\Psibold)\  \propto \  p_{\thetabold}(\Psibold|\Phibold_0)p_{\thetabold}(\Phibold_0).
	\label{eqn:bayesrule1}
\end{equation}
In this context, we classify conditioning into two categories (a) scenarios in which $p_{\thetabold}(\Psibold|\Phibold_0)$ is explicitly differentiable; (b) scenarios in which the relationship between generated state $\Phibold$ and condition vector $\Psibold$ is not differentiable and thus $p_{\thetabold}(\Psibold|\Phibold_0)$ cannot be explicitly obtained. Namely, for category (a), there exists a clear functional relationship between $\Phibold_0$ and $\Psibold$, whereas for category (b), such a relationship is absent. To address these two situations, we propose two different conditioning methods, (a) gradient-based conditional sampling without retraining and (b) conditional generation necessitating retraining, as described in the subsequent sections.

\subsubsection{Gradient-based conditonal generation without retraining}
\label{sec:gradient_method}
In the first scenario, the turbulent state variable is directly correlated with the condition vector through the relation $\Psibold = \mathcal{F}(\Phibold_0) + \epsilonbold_c$, where $\Fcal:\Rbb^{N_\mathrm{length}\times N_\mathrm{dof}}\rightarrow \Rbb^{N_\Psibold}$ is a functional mapping from spatiotemporal turbulent field $\Phibold_0$ to its associated condition vector $\Psibold$ (e.g., partial observation derived from $\Phibold_0$); $\epsilonbold_c$ represents zero-mean Gaussian noises with variance $\sigma_c^2$. By leveraging the Bayes' theorem, all the latent states conditioned on the vector $\Psibold$ can be expressed as,
\begin{equation}
	p(\Phibold_i|\Psibold) = \frac{p(\Psibold|\Phibold_i)p(\Phibold_i)}{p(\Psibold)},
	\label{eqn:bayesrule}
\end{equation}
where $p(\Psibold)$ is a normalizing constant that is intractable to compute in general. However, when using the score functions in terms conditional probability, the troublesome denominator is eliminated. Namely, we have,
\begin{equation}
	\nabla_{\Phibold_i} \log p(\Phibold_i|\Psibold) = \nabla_{\Phibold_i} \log p(\Psibold|\Phibold_i) + \nabla_{\Phibold_i} \log p(\Phibold_i),
	\label{eqn:gradient_equality}
\end{equation}
where the second term $\nabla_{\Phibold_i} \log p(\Phibold_i)$ can be obtained by the pre-trained score function $s_{\thetabold^*}(\Phibold_i; i)$ given by Eq.~\eqref{eqn:s_theta_star}. However, as there is not explicit dependency between $\Psibold$ and $\Phibold_i$, the first term $\nabla_{\Phibold_i} \log p(\Psibold|\Phibold_i)$ (gradient of log-likelihood term) needs to be approximated based on the relationship between $\Psibold$ and $\Phibold_0$. To this end, we factorize $p(\Psibold|\Phibold_i)$ as follows,
\begin{equation}
	\begin{split}
		p(\Psibold|\Phibold_i) = & \int p(\Psibold, \Phibold_0|\Phibold_i) d\Phibold_0 = \int p(\Psibold|\Phibold_0,\Phibold_i)p(\Phibold_0|\Phibold_i)d\Phibold_0 \\
		=& \int p(\Psibold|\Phibold_0)p(\Phibold_0|\Phibold_i)d\Phibold_0 =
		\Ebb_{\Phibold_0\sim p(\Phibold_0|\Phibold_i)}\big[p(\Psibold|\Phibold_0)\big],
	\end{split}
\end{equation}
which can be approximated as~\cite{chung2022diffusion}, 
\begin{equation}
	\Ebb_{\Phibold_0\sim p(\Phibold_0|\Phibold_i)}\Big[p(\Psibold|\Phibold_0)\Big] \approx p\Big(\Psibold|\Ebb_{\Phibold_0\sim p(\Phibold_0|\Phibold_i)}[\Phibold_0]\Big).
	\label{eqn:jensen_approx}
\end{equation}
The approximation error is theoretically bounded based on Jensen's inequality, known as the Jensen gap~\cite{gao2017bounds}. Now the conditional density $p(\Psibold|\Phibold_i)$ can be approximated by $p(\Psibold|\Ebb_{\Phibold_0\sim p(\Phibold_0|\Phibold_i)}[\Phibold_0])$, where $\Ebb_{\Phibold_0\sim p(\Phibold_0|\Phibold_i)}[\Phibold_0]$ is the pretrained diffusion model $\hat{\Phibold}_{\thetabold^*}(\Phibold_i;i)$ with optimized parameter $\thetabold^*$ (see Eq.~\eqref{eqn:phihati}). Accordingly, the gradient of the log likelihood in Eq.~\ref{eqn:gradient_equality} can be approximated as,
 \begin{equation}
	\nabla_{\Phibold_i} \log p(\Psibold|\Phibold_i) \approx \nabla_{\Phibold_i} \log p(\Psibold|\hat{\Phibold}_{\thetabold^*}(\Phibold_i;i)),
 	\label{eqn:grad_p_psi_phi}
 \end{equation}
 Based on the direct relationship between $\Psibold$ and $\Phibold_0$ as previously outlined, the conditional probability can be represented as 
 \begin{equation}
	p(\Psibold|\Phibold_{0}) \approx p(\Psibold|\hat{\Phibold}_{\thetabold^*}(\Phibold_i;i)) \sim \mathcal{N}\Big(\mathcal{F}\big(\hat{\Phibold}_{\thetabold^*}(\Phibold_i;i)\big), \sigma_c^2\Ibm\Big)
\end{equation}
By differentiating $\log p(\Psibold|\Phibold_{i})$ with respect to $\Phibold_i$ using the approximation in Eq.~\ref{eqn:grad_p_psi_phi}, we have
 \begin{equation}
	\nabla_{\Phibold_i} \log p(\Psibold|\Phibold_i) \approx -\frac{1}{\sigma_c^2} \frac{\partial \| \Fcal(\hat{\Phibold}_{\thetabold^*}(\Phibold_i;i))-\Psibold \|_{L2}^2}{\partial \Phibold_{i}},
	\label{eqn:logp}
\end{equation}
which can be obtained using the automatic differentiation capability of the neural networks. Namely,
\begin{equation}
	\pder{\Fcal(\hat{\Phibold}_{\thetabold^*}(\Phibold_i;i))}{\Phibold_i}=\pder{\Fcal(\hat{\Phibold}_{\thetabold^*}(\Phibold_i;i))}{\hat{\Phibold}_{\thetabold^*}(\Phibold_i;i)}\pder{\hat{\Phibold}_{\thetabold^*}(\Phibold_i;i)}{\Phibold_i}.
\end{equation}
As such, Eq.~\eqref{eqn:gradient_equality} can be computed as,
\begin{equation}
	\begin{split}
	\nabla_{\Phibold_i} \log p(\Phibold_i|\Psibold) \approx  \nabla_{\Phibold_i} \log p_{\thetabold^*}(\Psibold|\Phibold_i) + \nabla_{\Phibold_i} \log p_{\thetabold^*}(\Phibold_i) \\
	= - \frac{1}{\sigma_c^2} \nabla_{\Phibold_i} \big\| \Fcal(\hat{\Phibold}_{\thetabold^*}(\Phibold_i;i))-\Psibold \big\|_{L2}^2  + s_{\thetabold^*}(\Phibold_i; i),
	\end{split}
	\label{eqn:cond_scorefunc}
\end{equation}
For conditional generation, the recursive  unconditional sampling process as defined by Eq.~\ref{eqn:score_modeling} is guided by the gradient of the log likelihood term $\nabla_{\Phibold_i} \log p_{\thetabold^*}(\Psibold|\Phibold_i)$, which can be modified as,
\begin{equation}
	\tilde{\mubold}_{\thetabold^*}^{\mathrm{guide}}(\Phibold_{i}; \Psibold, i) = \frac{1}{\alpha_{i|i-1}}\Phibold_i +  \frac{\sigma^2_{i|i-1}}{\alpha_{i|i-1}}\Big(\sbm_{\thetabold^*}(\Phibold_{i}; i) - \omega \frac{1}{\sigma_c^2} \nabla_{\Phibold_i} \big\| \Fcal(\hat{\Phibold}_{\thetabold^*}(\Phibold_i;i))-\Psibold \big\|_{L2}^2 \Big),
	\label{eqn:cond_score_modeling}
\end{equation}
where $\omega$ represent a weighting parameter to balance the contributions from both gradient direction as we found that naively adding up both terms may cause low-quality generation with instability. In our implementation, we first compute the unit vector of the conditional sampling direction,
\begin{equation}
	\nbm^{\mathrm{guide}}_{\thetabold^*}(\Phibold_{i};\Psibold,i) \coloneqq \frac{\nabla_{\Phibold_i} \log p_{\thetabold^*}(\Psibold|\Phibold_i)}{\|\nabla_{\Phibold_i} \log p_{\thetabold^*}(\Psibold|\Phibold_i)\|_{L2}},
\end{equation}
and then the conditional sampling in terms of score functions is given as,
\begin{equation}
	\begin{split}
		\tilde{\mubold}^\mathrm{guide}_{\thetabold^*}(\Phibold_{i};i-1,i) =& \mubold_{\thetabold^*}(\Phibold_{i};i-1,i) + \\
		&\beta_i^\mathrm{guide}\max\Bigl(
		||\frac{\sigma^2_{i|i-1}}{\alpha_{i|i-1}}\sbm_{\thetabold^*}(\Phibold_{i};i)||_2,
		(\sigma^\mathrm{guide}_i)^2
		\Bigr) 
		\nbm^{\mathrm{guide}}_{\thetabold^*}\Big(\Phibold_{i};\Psibold,i\Big),
	\end{split}
	\label{eqn:cond}
\end{equation}
where $\beta_i^\mathrm{guide}$ and $\sigma^\mathrm{guide}_i$ are hyper-parameters, whose magnitudes should be kept relatively small based our empirical observations. To further enhance the results, we can iterate the conditional gradient updating in the last denoising step for $N_\mathrm{refine}$ times.

\begin{remark}
Gradient-based conditional sampling does not need to retrain the diffusion model. Namely, the diffusion model can be trained unconditionally. Once trained,  it enables the generation of new turbulent flows under various conditions, as indicated by Eq.~\ref{eqn:cond}.
\end{remark}

\begin{remark}
Maintaining a balance between the two directional terms in Eq.~\ref{eqn:cond_scorefunc} is important. An excessively large gradient of the conditional probability density may result in samples that meet specific conditions but situated in low-density regions of the unconditional distribution, yielding suboptimal samples. Our objective is to generate samples that not only reside within the dense regions of the unconditional distribution but also align with the set conditions. Considering the challenges associated with direct density evaluation in the diffusion model, it is advisable to accentuate the leading gradient of the unconditional density, while leverage the gradient of the conditional density to guide the sampling with a relatively smaller weight. 
\end{remark}

\subsubsection{Concatenation-based conditional generation through retraining}
\label{sec:conditionwithconcatanation}
In this subsection, we address the second conditioning scenario, where there isn't a clear analytical or differentiable relationship between the condition vectors $\Psibold$ and the turbulent flow solutions $\Phibold_0$. Take, for instance, the generation of eddy-resolved LES turbulence conditioned on its counterpart URANS simulation results. When training the diffusion model on LES datasets, the relationship between URANS and LES outcomes hinges on shared geometry, Reynolds number, and the fact that both models arise from identical governing equations, albeit with varying modeling assumptions. However, the direct mapping function $\mathcal{F}: \Psibold \to \Phibold_0$ and its associated gradients, which would establish a tangible link between RANS and LES results, are elusive, given the high-dimensional intricacies (space-time degrees of freedom), solver inconsistencies (as they may originate from distinct solvers), and the considerable computational demands.  Such complexities make the gradient-based conditional generation technique inapplicable in this context.

As a viable alternative strategy, conditions can be incorporated into the  neural networks' input, necessitating a retraining process with conditioned samples to implicitly approximate the true conditional distribution. More specifically, the U-Net's input is modified by concatenating the condition vector $\Psibold$ with the perturbed hidden state $\Phibold_{i}$. The training procedure remains the same as that for the unconditional diffusion model as discussed in Section~\ref{sec:VDF}, where the Gaussian transition kernel is unaltered. In this way, the conditional mapping $\mathcal{F}: \Psibold \to \Phibold_0$ is implicitly learned by the neural networks, and thereby directing the flow generation conditioned on $\Psibold$.  Tables~\ref{tab:traning} and~\ref{tab:sampling} provide a comparison between the training and sampling processes of the unconditional and conditional diffusion models, respectively. 
\begin{table}[H]
\begin{center}
			\begin{tabular}{ |c|c| } 
					\hline
					&Training optimization\\ 
					Unconditional & $\thetabold^* =	\argoptunc{\thetabold}{\sum_{\Phibold_{0}\in\Acal_\mathrm{train}}\Ebb_{\epsilonbold\sim \Ncal(\mathbf{0},\Ibm), i\sim\Ucal(1,N_\mathrm{noise})}\big[C_i\|\hat{\Phibold}_{\thetabold}(\Phibold_i;i)-\Phibold_0\|^2_{L2}\big]}$ \\ 
					Conditional  &$\thetabold^\circ =	\argoptunc{\thetabold}{\sum_{\Phibold_{0}\in\Acal_\mathrm{train}}\Ebb_{\epsilonbold\sim \Ncal(\mathbf{0},\Ibm), i\sim\Ucal(1,N_\mathrm{noise})}\big[C_i\|\hat{\Phibold}_{\thetabold}(\Phibold_i;i,\Psibold)-\Phibold_0\|^2_{L2}\big]}$ \\ 
					\hline
				\end{tabular}
		\end{center}
\caption{A comparison between the training of unconditional and conditional diffusion models.}
\label{tab:traning}
\end{table}
\begin{table}[H]
	\begin{center}
			\begin{tabular}{ |c|c|c|c|c| } 
					\hline
					&Generated state & Mean of reverse kernel & Noise & Score function\\ 
					Unconditional&$\hat{\Phibold}_{\thetabold^*}(\Phibold_i;i)$ & $\tilde{\mubold}_{\thetabold^*}(\Phibold_{i};i-1,i)$ & $\epsilonbold_{\thetabold^*}(\Phibold_i;i)$& $\sbm_{\thetabold^*}(\Phibold_i;i)$\\ 
					Conditional&$\hat{\Phibold}_{\thetabold^\circ}(\Phibold_i;i,\Psibold)$ & $\tilde{\mubold}_{\thetabold^\circ}(\Phibold_{i};i-1,i,\Psibold)$ & $\epsilonbold_{\thetabold^\circ}(\Phibold_i;i,\Psibold)$& $\sbm_{\thetabold^\circ}(\Phibold_i;i,\Psibold)$\\ 
					\hline
				\end{tabular}
		\end{center}
\caption{A comparison between the sampling of unconditional and conditional diffusion models.}
\label{tab:sampling}
\end{table}

\subsection{Generating long-span spatiotemporal turbulence}
\label{sec:longvideo}
The proposed diffusion model is adept at generating spatiotemporal turbulence fields $\Phibold_0\in\Rbb^{N_\mathrm{length} \times N_\mathrm{dof}}$, consisting of a fixed set of $N_\mathrm{length}$ temporal snapshots. Due to the intrinsic limitations imposed by memory footprints, $N_\mathrm{length}$ cannot be substantially large during the training phase. Nonetheless, for enhanced applicability in real-world scenarios, there is an imperative need to extrapolate the capabilities of the trained diffusion model to enable the synthesis of turbulent sequences that span duration considerably exceeding the default length $N_\mathrm{length}$. To this end, we propose an autoregressive conditioning sampling strategy, which allows us to robustly generate long-span turbulent flows with arbitrary temporal length. Specifically, the spatiotemporal turbulent state $\Phibold_0$ of default length is decomposed into two non-overlapping subsequences: 
\begin{equation}
	\Phibold_{0} = [\Phibold^a_{0},\Phibold^b_{0}],
\end{equation}
where $\Phibold^a_{0}\in\Rbb^{N_\mathrm{previous} \times N_\mathrm{dof}}$ represents the preceding subsequence, whereas $\Phibold^b_{0}\in\Rbb^{(N_\mathrm{length} - N_\mathrm{previous} )\times N_\mathrm{dof}}$ denotes the subsequent flow subsequence. Now the autoregressive conditional generation is formulated as follows: given an preceding flow sequence $\Phibold_0^a$, the goal is to generate a subsequent flow sequence $\Phibold_0^b$ based on the conditional probability density $p(\Phibold_0^b|\Phibold_0^a)$. Once the diffusion model is trained with optimal parameters $\thetabold^*$, the subsequent flow sequence can be obtained via conditional sampling,
\begin{equation}
	\quad\Phibold^b_{0} \sim p_{\thetabold^*}(\Phibold^b_{0}|\Phibold^a_{0}),
	\label{eqn:condition_long_video}
\end{equation}
By repetitively sampling from these conditional probability distributions in an autoregressive manner, our approach holds the capability to synthesize turbulent sequences with a arbitrary length. Significantly, this can be achieved \emph {without the need of retraining the model}. 

\subsubsection{Gradient-based autoregressive conditioning method}
The gradient-based conditional generation method, introduced in Section \ref{sec:gradient_method}, can be used to achieve this goal. In this context, the conditional vector is the preceding subsequence, $\Psibold \coloneqq \Phibold_{0}^a$. Within the gradient-based conditional generation framework, the gradient of log likelihood associated with the preceding subsequence is used to guide the generation of the subsequent sequence $\Phibold_{0}^b$. Specifically, we autoregressively sample the conditional distribution $p(\Phibold_{0} | \Phibold_{0}^{a})$ of the turbulence sequence $\Phibold_0$ with the default length $N_{\mathrm{length}}$ given the condition vector, i.e., $\Phibold_{0}^a$. The state-to-condition mapping $\mathcal{F}$ is defined as,
\begin{equation}
	\Fcal: \Rbb^{N_\mathrm{length}\times N_\mathrm{dof}} \to \Rbb^{\frac{1}{2}*N_\mathrm{length}\times N_\mathrm{dof}},
\end{equation}
which selected out the first $\frac{1}{2}*N_\mathrm{length}$ snapshots of the generated flow sequence $\hat{\Phibold}_{\thetabold^*}$, i.e., 
\begin{equation}
	\Fcal(\hat{\Phibold}_{\thetabold^*}) = \hat{\Phibold}^a_{\thetabold^*},
\end{equation}
where $\hat{\Phibold}^a_{\thetabold^*}$ is the first half portion of the full sequence $\hat{\Phibold}_{\thetabold^*}$. The key for conditional sampling is to approximate the gradient of log likelihood of condition vector $\nabla_{\Phibold_i} \log p(\Psibold|\Phibold_i)$, which involve the derivative computation as follows,
 \begin{equation}
	 \frac{\partial \| \Fcal(\hat{\Phibold}_{\thetabold^*}(\Phibold_i;i))-\Psibold \|_{L2}^2}{\partial \Phibold_{i}} = \pder{||\hat{\Phibold}^a_{\thetabold^*}(\Phibold_i;i)-\Phibold^a_0||_2^2}{\Phibold_i}.
\end{equation}
Then, the subsequent operations are detailed in Section~\ref{sec:gradient_method}. By adopting this autoregressive approach, the model can successively generate segments of the flow sequence. Each subsequent segment is influenced by the preceding segment, ensuring smooth and coherent transitions in the turbulent dynamics. This iterative procedure, when executed repeatedly, facilitates the generation of flow sequences of any desired length. This methodology offers a potent and versatile solution for extending the capabilities of our trained diffusion model beyond its inherent limitations, fulfilling the demand for long-duration turbulent sequences in various practical applications.

\subsubsection{Replacement-based autoregressive method}
\label{sec:longvideo_repalce}
An alternative autoregressive method has been discussed by Ho et al.~\cite{ho2022video}, targeting the generation of extended videos. While the foundational approach remains rooted in unconditional sampling, the strategy involves modifying partial hidden states to enable autoregressive generation.
Specifically, the (hidden) state $\Phibold_{i}$ is partitioned as $\Phibold_{i} = [\Phibold_{i}^a, \Phibold_{i}^b]$. We aim to compute the expectation of $\Phibold_{0}^b$ over the conditioned probability $p_\theta^*(\Phibold_{0}^b | \Phibold_{0}^a, \Phibold_{i}) = p_\theta^*(\Phibold_{0}^b | \Phibold_{0}^a, [\Phibold_{i}^a, \Phibold_{i}^b])$ as,
\begin{equation}
	\begin{split}
		\Ebb_{\Phibold^b_0\sim p_{\thetabold^*}(\Phibold^b_{0} | \Phibold_i)}\Big[\Phibold^b_0\Big]
		&= \int p_{\thetabold^*}(\Phibold^b_{0}|\Phibold^a_{0},[\Phibold^a_i,\Phibold^b_i])\Phibold^b_{0}d\Phibold^b_{0}\\
		&=\int \frac{p_{\thetabold^*}(\Phibold^b_{0}|\Phibold^a_{0},[\Phibold^a_i,\Phibold^b_i]) p_{\thetabold^*}(\Phibold^b_{0}| [\Phibold^a_{0},\Phibold^b_i])}{p_{\thetabold^*}(\Phibold^b_{0}| [\Phibold^a_{0},\Phibold^b_i] )}\Phibold^b_{0}d\Phibold^b\\
		&=\Ebb_{\Phibold^b_0\sim p_{\thetabold^*}(\Phibold^b_{0}| [\Phibold^a_{0},\Phibold^b_i] )}\left[\Phibold^b_0  \frac{p_{\thetabold^*}(\Phibold^b_{0}|\Phibold^a_{0},[\Phibold^a_i,\Phibold^b_i])}{p_{\thetabold^*}(\Phibold^b_{0}| [\Phibold^a_{0},\Phibold^b_i] )} \right].
		\label{eqn:Ebright}
	\end{split}
\end{equation}
By assuming $p_{\thetabold^*}(\Phibold^b_{0}|\Phibold^a_{0},[\Phibold^a_i,\Phibold^b_i]) \approx p_{\thetabold^*}(\Phibold^b_{0}| [\Phibold^a_{0},\Phibold^b_i] )$ as iterative denoised $\Phibold_i^a$ converges to $\Phibold_0^a$, we have
\begin{equation}
	\begin{split}
		\Ebb_{\Phibold^b_0\sim p_{\thetabold^*}(\Phibold^b_{0}| [\Phibold^a_{0},\Phibold^b_i)]}\left[\Phibold^b_0\frac{p_{\thetabold^*}(\Phibold^b_{0}|\Phibold^a_{0},[\Phibold^a_i,\Phibold^b_i])}{p_{\thetabold^*}(\Phibold^b_{0}| [\Phibold^a_{0},\Phibold^b_i] )} \right] &\approx \Ebb_{\Phibold^b_0\sim p_{\thetabold^*}(\Phibold^b_{0}|[\Phibold^a_{0},\Phibold^b_i])}\Big[\Phibold^b_0\Big]\\
		&=\hat{\Phibold}^b_{\thetabold^*}([\Phibold^a_{0},\Phibold^b_i];i)
	\end{split}
	\label{eqn:replacement}
\end{equation}
where $\hat{\Phibold}^b_{\thetabold^*}([\Phibold^a_{0};\Phibold^b_i];i)$ is from the partition of, 
\begin{equation}
	\hat{\Phibold}_{\thetabold^*}([\Phibold^a_{0};\Phibold^b_i];i)= \big[\hat{\Phibold}^a_{\thetabold^*}([\Phibold^a_{0};\Phibold^b_i];i);\hat{\Phibold}^b_{\thetabold^*}([\Phibold^a_{0};\Phibold^b_i];i)\big].
\end{equation}
Namely, the core of the replacement method is to replace the first half of the hidden vector $\Phibold_{i}^a$ with the given $\Phibold_{0}^a$ during unconditional sampling process. 

\begin{remark}
The formulation in in \eqref{eqn:replacement} operates under the assumption $p_{\thetabold^*}(\Phibold^b_{0}|\Phibold^a_{0},[\Phibold^a_i,\Phibold^b_i]) \approx p_{\thetabold^*}(\Phibold^b_{0}| [\Phibold^a_{0},\Phibold^b_i] )$. The validity of this assumption in the context of estimating the density value ratio (which corresponds to the importance sampling weight in \eqref{eqn:Ebright}) remains unclear, even with an adequately trained model, i.e., $i \to 0$. Since $p_{\thetabold^*}(\Phibold^b_{0}|\Phibold^a_{0},[\Phibold^a_i,\Phibold^b_i])$ is always greater than $p_{\thetabold^*}(\Phibold^b_{0}| [\Phibold^a_{0},\Phibold^b_i] )$ due to more information in the conditioning, the density ratio is always greater then one. The replacement method's disregard for the importance weights exceeding unity might lead to underestimated values. As sequences are progressively generated, this diminishing effect could compound, aligning with our observations in the experimental study presented in Section \ref{sec:pitz}.
\end{remark}

\section{Numerical results and discussions}
\label{sec:results}

\newcommand{\rectangle}{\raisebox{0pt}{\tikz{\draw[purple,solid,line width = 1.0pt,fill=purple](2.mm,0) rectangle (3.5mm,1.5mm);\draw[-,purple,solid,line width = 1.0pt](0.,0.8mm) -- (5.5mm,0.8mm)}}}
\newcommand{\rectangleblue}{\raisebox{0pt}{\tikz{\draw[cyan,solid,line width = 1.0pt,fill=cyan](2.mm,0) rectangle (3.5mm,1.5mm);\draw[-,cyan,solid,line width = 1.0pt](0.,0.8mm) -- (5.5mm,0.8mm)}}}
\newcommand{\graycircle}{\raisebox{0pt}{\tikz{\draw[gray,solid,line width = 1.0pt,fill=gray](2.7mm,0) circle (0.8mm);\draw[-,gray,solid,line width = 1.0pt](0.,0.0mm) -- (5.5mm,0.0mm)}}}

To showcase the robust and versatile turbulence generation capabilities of the proposed conditional diffusion model, we conducted on a series of numerical experiments, exploring various generation scenarios on three distinct turbulent flow cases:
\begin{enumerate}
	\item 2D unsteady flows over a backward-facing step: In this case, we highlight the model's capability in generating LES-like instantaneous eddy-resolved turbulent flows when provided with URANS simulated flow solutions.
	\item 3D turbulent channel flows: Here, the diffusion model is trained to adeptly generate instantaneous spatiotemporal sequences of turbulent channel flow, given specified initial conditons, statistics, or entirely from scratch.
	\item 3D compressible supersonic turbulent boundary layers: This case underscores the model's super-resolution generation capabilities - where the high-resolution DNS high-speed turbulent boundary layers are generated conditioned on low-resolution input measurements.
\end{enumerate}

\subsection{RANS-conditioned generation of eddy-resolved turbulence over backward-facing step}
\label{sec:pitz}

In this section, we showcase the capabilities of our proposed method in generating LES-like spatiotemporal flow realizations, predicted upon a URANS-simulated flow. We further compare the performance of different variants of our proposed method. Consider a 2D channel featuring a backward-facing step; the specific configuration of this flow case is depicted in Fig.~\ref{fig:pitz_geo}. Our focus revolves around the unsteady velocity, as a spatiotemporal vector field $\mathbf{u}(x,y,t) = [u_1(x, y, t), u_2(x, y, t)]^T :\Omega\times\Rbb_{\geq0}\rightarrow\Rbb^2$, where $\Omega$ represents the computational domain. A key area of our interest is the flow separation zone after the step, indicated as the shaded area in Fig.~\ref{fig:pitz_geo}. In LES simulations, introducing random perturbations to the inlet yields multiple realizations of unsteady flow sequences even at the same Reynolds number. In contrast, the URANS simulation inherently averages out these stochastic fluctuations. Consequently, for a specified Reynolds number, every distinct LES realization of instantaneous flow sequence can be associated with a single URANS simulated flow sequence.  Essentially, these LES flow realizations can be interpreted as samples drawn independently from a stochastic distribution conditioned on the URANS result for a given Reynolds number. The LES and URANS simulation details are provided in Tab.~\ref{table:les_rans_data}.

\subsubsection{Data preparation and model training}

The diffusion model is developed to generate eddy-resolved instantaneous flow sequences given different Reynolds number. To produce the flow dataset, we consider a parameter set $\Dcal$ consisting of $11$  Reynolds numbers ($Re$) evenly distributed between $5000$ to $14000$,
\begin{equation}
\Dcal=\{Re_i = 5000+(i-1)\times900 | i=1, \cdots,11 \}.
\end{equation}
At each Reynolds number ($Re_i$), we perform one URANS simulation and the simulated unsteady flow sequence ($\Psibold$) contains 240 snapshots 
\begin{equation}
\Psibold^{Re_i} = [ \psibold^{Re_i}_1,\cdots \psibold^{Re_i}_{240} ], \quad i = 1, \cdots, 11,
\end{equation}
and LES simulations with five realizations of random inlet perturbations ($\Phibold$)
\begin{equation}
\Phibold^{Re_i}_k = [ \phibold^{Re_i}_{k,1},\cdots \phibold^{Re_i}_{k,240} ], \quad i = 1, \cdots, 11, k=1,\cdots , 5.
\end{equation} 
Subsequently, we partition the parameter set into two subsets:, the testing set ($\Dcal_\mathrm{test}$) and the training set ($\Dcal_\mathrm{train}$),
\begin{equation}
	\Dcal_\mathrm{test} = \{5900,13100\},\quad \Dcal_\mathrm{train} = \Dcal\setminus\Dcal_\mathrm{test}.
\end{equation}
Considering memory constraints and the need for data augmentation, the complete flow sequence of 240 snapshots is divided into 200 shorter subsequences, each containing 40 snapshots, for training purposes. In testing phase, long flow sequence can be generated using auto-regressive conditional sampling as discussed above. Specifically, the training sets for condition vectors and corresponding flow solutions are given as,
\begin{equation}
\begin{split}
	&\Vcal = \big\{ \{\psibold^{Re_i}_j,\dots\psibold^{Re_i}_{j+40}\} \big| Re_i\in\Dcal_\mathrm{train},j=1,\dots,200 \big\},\\
	&\Wcal = \big\{ \{\phibold^{Re_i}_{k,j},\dots\phibold^{Re_i}_{k,j+40}\} \big| Re_i\in\Dcal_\mathrm{train},j=1,\dots,200,k=1,\dots, 5 \big\}.\\
\end{split}
\label{eqn:data_aug}
\end{equation}
\begin{remark}
 We consistently adopt the data augmentation strategy outlined in \eqref{eqn:data_aug} to address the challenges posed by limited training data. This approach offers an advantage in generating a greater number of sequences compared to simply dividing the lengthy sequence into discrete non-overlapping segments.
\end{remark}

\subsubsection{URANS-conditioned generation of LES flow sequences on testing $Re$}

\begin{figure}[H] 
	\centering
	
	\includegraphics[width=0.9\textwidth]{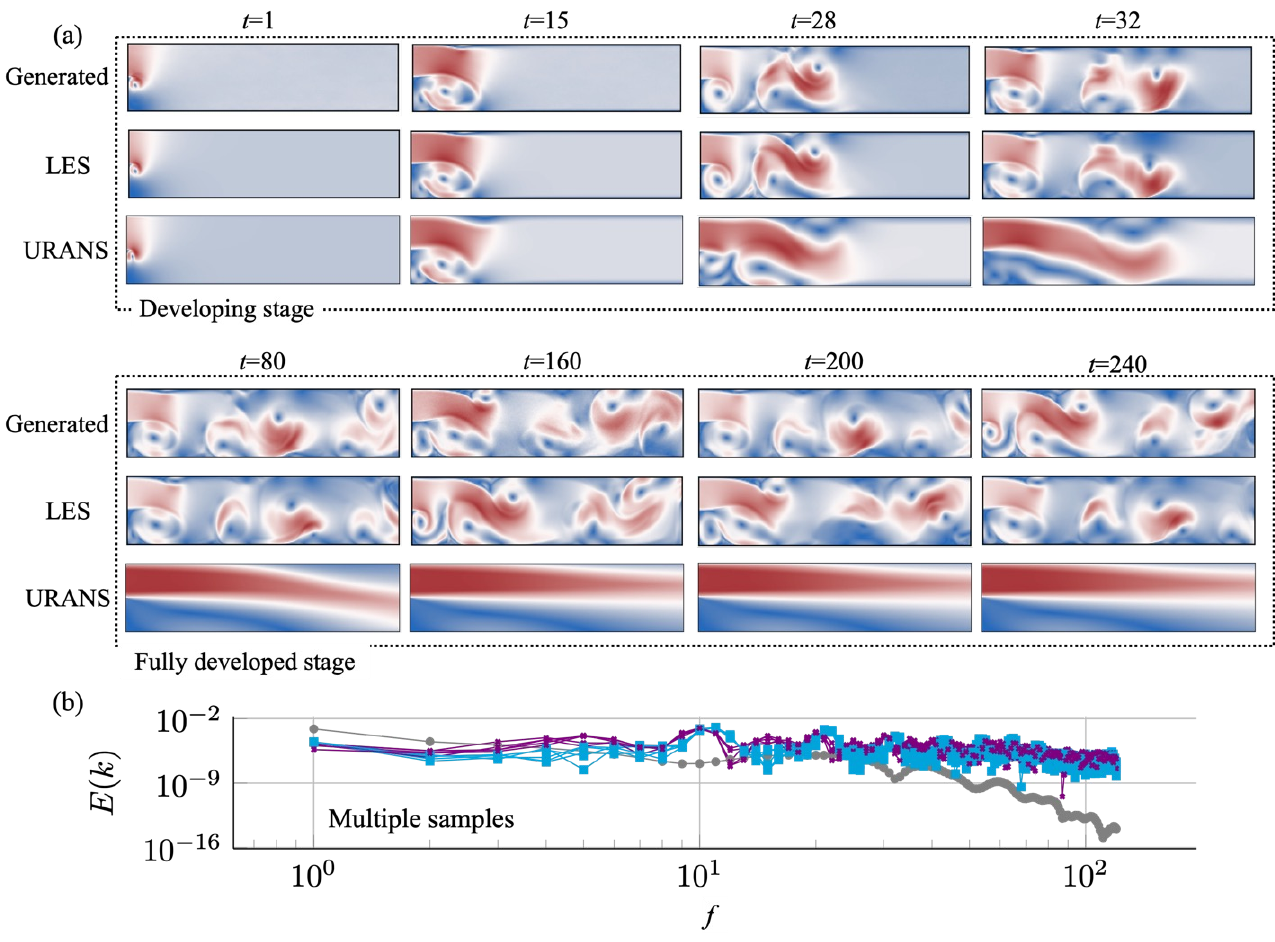} 
	\caption{(a) Velocity magnitude of one generated samples, LES  and URANS at $Re=5900$ at developing and fully developed stages. (b) the energy spectra for multiple generated samples (\protect\rectangleblue), compared with LES (\protect\rectangle) and URANS (\protect\graycircle)  at $x=4, y=1$. }
	\label{fig:pitz:vmag_Re5900}
\end{figure}

Given a testing Reynolds number (e.g., $Re = 5900$), URANS is conducted to serve as the conditioning for the LES-like generation. As shown in Fig.~\ref{fig:pitz:vmag_Re5900}, a long LES-like turbulent flow over a backward-facing step is generated by the gradient-based autoregressive method. We first focus on the flow development phase (first 40 steps), specifically observing the formation and growth of the recirculation bubble as the flow progresses past the step. For one realization of the generated flow sequence (see contours in Fig.~\ref{fig:pitz:vmag_Re5900}),  it is noticeable that the generated sequence retains similarities with the URANS simulated one up to approximately the 15th timestep, with significant divergences emerging after the 20th timestep.  Post the 40th timestep, although the URANS-simulated conditions tend to stabilize, the generated samples retain their turbulent nature, marked by the dynamically changed vortex structures, unaffected by the nearly-steady URANS conditioning. This observation suggests that our model actually generates new flow features, advancing beyond mere replication or upscaling of URANS data, which is fundamentally different from previous SR works~\cite{geneva2020multi}. By comparing the contours of one realization of generated flow sequences with one LES sequence, we found the unsteady flow patterns, vortex structures and level of details are very similar. It is important to note that the instantaneous flow fields between the generated and actual LES samples inherently differ, as they are independent random realizations and lack any direct comparison.

To further compare our generative model with LES statistically, we generate multiple realizations based on the same URANS simulation. The energy spectra of the realizations from both LES and diffusion models behind the step ($x=4, y=1$) are plotted in Fig.~\ref{fig:pitz:vmag_Re5900}, where the URANS result is also shown for comparision. As observed in Fig.~\ref{fig:pitz:vmag_Re5900}, the URANS simulation's energy spectrum decays rapidly at high frequencies and notably lacks the pronounced peak at $f=10$. This absence of a peak in the URANS spectrum is indicative of the model's limited capacity to resolve the eddies and capture the recirculation region in the wake after the step.  In contrast, all realizations from our diffusion model effectively generate these eddies and the recirculation regions and their alignment with the LES results is evident, as their energy spectra show a consistent peak value and statistical characteristics. Additionally, the energy spectra for different realizations exhibit slight differences, indicating the generation process in our model is non-deterministic. Similar conclusions are drawn for a higher testing Reynolds number ($Re=13100$), as evidenced in Fig.~\ref{fig:pitz:vmag_Re13100}. This demonstrates the capability of our Bayesian-based diffusion model to produce diverse LES-like instantaneous eddy-resolved turbulent flows when provided with URANS-simulated flow solutions for a specified $Re$.

We also examine the mean velocities and their fluctuations at various locations in the wake region to spatially assess the statistics of our generated samples. Figure~\ref{fig:pitz:umean_uvvar_Re5900} shows a comparison between the diffusion-generated results and the LES/URANS simulations for $Re=5900$ across nine representative locations, including recirculation, reattachment, and recovery areas. The profiles of the first and second moments of the generated samples agree with those of LES simulated results very well in both developing and fully developed stages. While URANS simulations reasonably capture the mean velocities and their fluctuations during the developing stage, they significantly underperform for the fully developed flows. Notably, URANS simulations substantially underestimate velocity fluctuations when the flow become fully developed, primarily due to their inability to accurately capture the recirculation region in the wake. This leads to notable discrepancies in mean velocities, especially in the region defined by $x=[0,6]$ and $y<1$. Beyond the recirculation region, particularly in the area of $x=[7,9]$ and $y<1$, URANS simulations align more closely with LES results. However, URANS tends to overestimate mean velocities in regions where $y>1$. Conversely, our diffusion model results can generate intricate vortices and flow patterns which align well with LES results, as evident by their agreements in velocity mean and fluctuation profiles across all of these wake regions. The same observation and conclusion can be obtained for the case of $Re=13100$, as shown in Fig.~\ref{fig:pitz:umean_Re13100}. 
\newcommand{\blue}{\raisebox{0.5pt}{\tikz{\node[draw,scale=0.4,circle,fill=blue!20!blue](){};}}}
\newcommand{\green}{\raisebox{0.5pt}{\tikz{\node[draw,scale=0.4,circle,fill=green!20!green](){};}}}
\newcommand{\magenta}{\raisebox{0.5pt}{\tikz{\node[draw,scale=0.4,circle,fill=magenta!20!magenta](){};}}}
\begin{figure}[tp]
	\centering
	\includegraphics[width=1.0\textwidth]{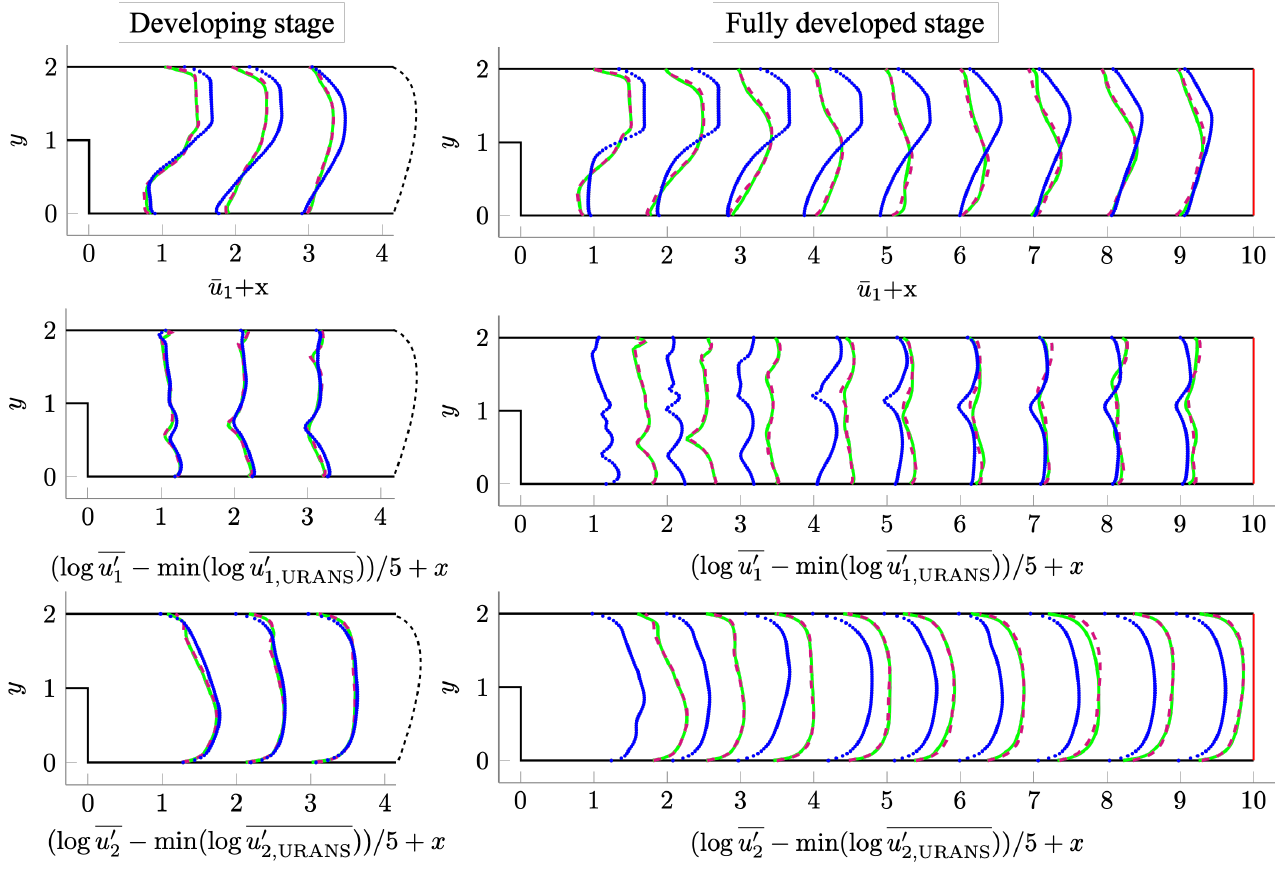} 
	\caption{Mean of streamwise velocity, variance of streamwise velocity and variance of wall-normal velocity at $Re = 5900$for developing and fully developed stages from LES (\protect\green), URANS (\protect\blue) and diffusion model (\protect\magenta) evaluated at $x = 0.1$ (\textit{left column}), $x = 0.2$ (\textit{middle column}) and $x = 0.3$ (\textit{right column}).}
	\label{fig:pitz:umean_uvvar_Re5900}
\end{figure}

\subsubsection{Comparison of different autoregresive sampling methods}

\begin{figure}[tp]
\centering
\input{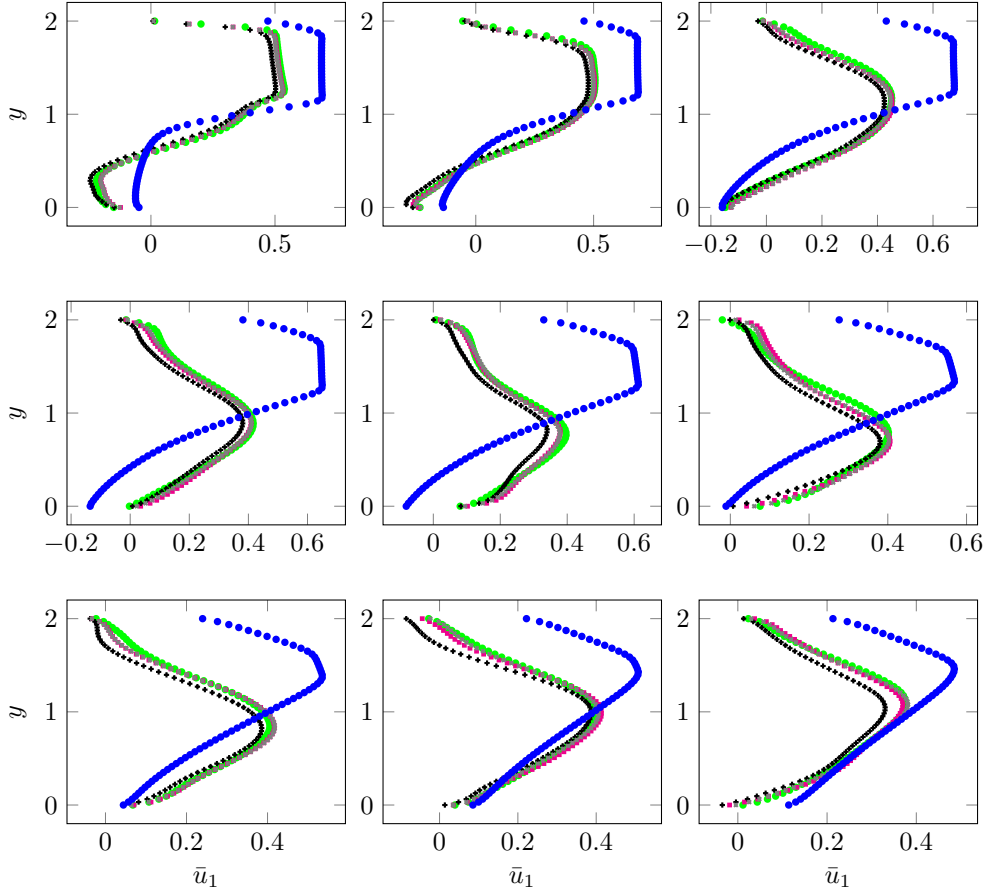}
\caption{Mean of streamwise velocity at $Re=13100$ in the fully-developed phase: LES (\ref{line:pitz:mean_les}), URANS (\ref{line:pitz:mean_rans}), Diff-Gradient (\ref{line:pitz:mean_diff}), Diff-Replace (\ref{line:pitz:mean_replace}) and Diff-Vanilla (\ref{line:pitz:mean_vanilla}). The figures, arranged from left to right and top to bottom, correspond to locations at $x=[1,9]$.
}
\label{fig:pitz:umean_Re13100_developed_compare_three_diff_method}
\end{figure}

We previously introduced gradient-based (Secion~\ref{sec:gradient_method}, referred to as Diff-Gradient) and replacement-based (Section~\ref{sec:conditionwithconcatanation}, referred to as Diff-Replace) autoregressive sampling method in details. To assess the performance of these methods, along with the original unconditional direct generation method proposed by Ho et al.\cite{ho2022video} (Section \ref{sec:unconditionalsample}, referred to as Diff-Vanilla), we conducted a comparative analysis of sample statistics at nine locations ($x=[1,9]$) for $Re=13100$.  As shown in Fig.\ref{fig:pitz:umean_Re13100_developed_compare_three_diff_method}, all three methods are capable of generating samples conditioned on URANS, effectively capturing the mean velocity within the recirculation region and far wakes. However, the replacement-based method slightly underestimates the mean velocity, particularly in the far wake region. Regarding long-span generation, the direct generation method is impractical due to memory constraints. Although both autoregressive generation methods can produce arbitrarily long sequences, the replacement-based method tends to exhibit a continuous decrease in velocity magnitude over time. This trend is attributed to noticeable energy dissipation during self-conditioning rollouts, as discussed in Section~\ref{sec:longvideo_repalce}. Our proposed gradient-based autoregressive generation method demonstrates robust capability in generating LES-like turbulence over extended durations

We also evaluated the computational cost of three sampling methods using the diffusion model, as shown in Fig.~\ref{fig:pitz:cost_compare}. Since the URANS computation is performed only once and serves as a condition, its cost is not included in the evaluation for the generation of multiple realizations. In contexts requiring numerous short and coherent sequences, such as those focusing exclusively on developed flows, the vanilla method (Section \ref{sec:unconditionalsample}) is the most efficient, offering significant speed improvements. Conversely, for generating long and coherent sequences, the gradient-based autoregressive method is preferable due to its superior performance and faster operation compared to the replacement-based autoregressive method. It's important to note that this initial example primarily demonstrates the model's ability to parametrically convert dynamics characterized by large-scale features into those exhibiting small-scale and chaotic features, while the speedup observed in this case due to the computational efficiency of the small-scale 2D LES simulation. It is anticipated that the benefits in terms of computational efficiency for the diffusion model will be more significant in scenarios involving large-scale and more intricate 3D turbulence simulations, which will be further studied in the subsequent sections.  



\subsection{Generation of instantaneous turbulent channel flows under various conditions}
\label{sec:channel}

We now demonstrate the efficacy of our proposed method in generating instantaneous DNS velocity fields of a 3D turbulent channel  flow form scratch or from specified conditions such as initial flow snapshot, RMS profiles of velocity fluctuations, and Reynolds stress. For the fully developed turbulent channel flow at $Re_\tau=180$, governed by the unsteady incompressible Navier-Stokes equations~\cite{kim1987turbulence}, the flow exhibits homogenetiy in the streamwise and spanwise directions. In this scenario, the turbulence are statistically homogeneous except in the wall-normal direction ($y^+$) \cite{pope2000turbulent}. This one-dimensional characteristic facilitates the synthesis of independent two-dimensional samples from uncorrelated sections perpendicular to the streamwise direction \cite{kim1987turbulence, moser1999direct, pope2000turbulent, yousif2022physics}. This one-dimensional characteristics allows us to synthesize independent two-dimensional instantaneous flow sequences from uncorrelated sections perpendicular to the streamwise directions \cite{kim1987turbulence,moser1999direct,pope2000turbulent,yousif2022physics}. Our focus is on generating the 3D velocity fields ($\mathbf{u}(y,z,t)=[u_1(y,z,t),u_2(y,z,t),u_3(y,z,t)]^T: \Omega\times\Rbb_{\geq0}\rightarrow\Rbb^2$) at the the channel cross-section. We will explore in detail the application of the gradient-based autoregressive conditional diffusion method for generating long-span, spatiotemporal dynamics of the 3D channel flows.


\subsubsection{Data preparation and model training}
The diffusion model is built to produce the the cross-sectional velocity field of a 3D channel of dimensions $L_x\times \L_y \times Lz=2\pi \times 2 \times \pi$ at $Re_{\tau}=180$,  with an emphasis on generating extended sequences under various conditions. The dataset is obtained through fully-resolved DNS runs, where the cross section is discretized by $N_y\times N_z=256\times128$. The instantaneous velocity fields of a total of 14 uncorrelated cross-sections along the streamwise direction, spanning 4 flow-through times, are collected to create our dataset. Each flow-through time includes 300 time steps, resulting in a total of 16,800 snapshots. Of these, 86\% are utilized for training, and the remaining 14\% serve as test set for turbulence statistics comparison. 

\subsubsection{Direct generation of long sequential turbulence from scratch}
\begin{figure}[tp]
	\includegraphics[width=0.9\textwidth]{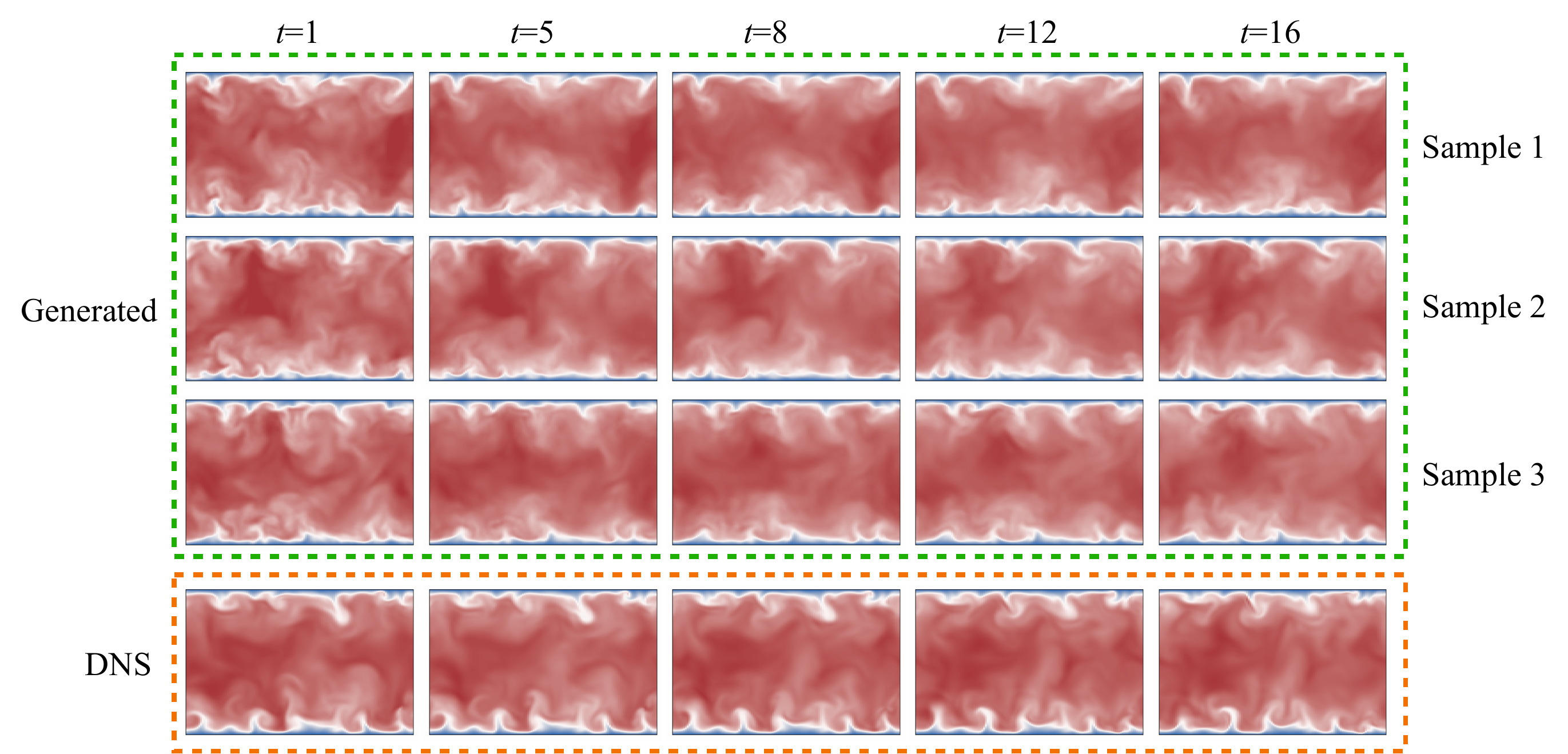}
	\caption{Spatiotemporal sequences of instantaneous streamwise velocity fields generated from our diffusion model and DNS.} 
	\label{fig:idns:ucontour_uncondition}
\end{figure}

\begin{figure}[tp]
	\centering
	\input{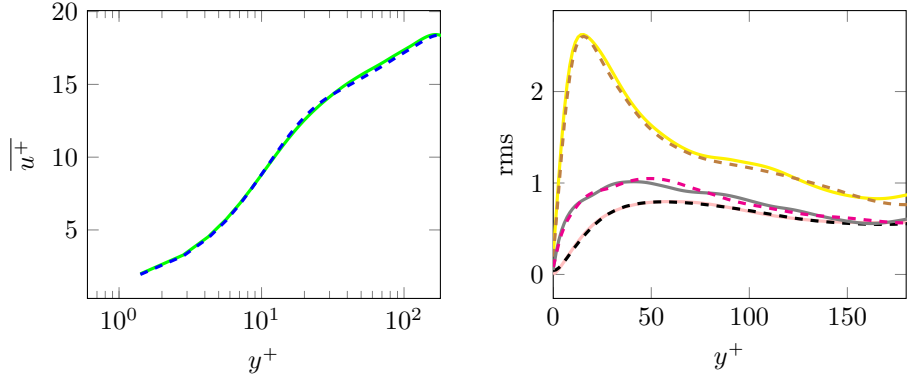}
	\caption{The mean streamwise velocity profile (\textit{left}) from DNS (\ref{line:idns:mean_u_dns}) and diffusion model (\ref{line:idns:mean_u_diff}); The RMS profiles (\textit{right}) of streamwise velocity fluctuation from DNS (\ref{line:idns:urms_dns}), diffusion model (\ref{line:idns:urms_diff}), of wall-normal velocity fluctuation from DNS (\ref{line:idns:vrms_dns}), diffusion model (\ref{line:idns:vrms_diff}) , and of spanwise velocity fluctuation from DNS (\ref{line:idns:wrms_dns}), diffusion model (\ref{line:idns:wrms_diff}). 
		\label{fig:idns:mean_and_fluctuation_uncondition}
	}
\end{figure}

We initially generated the flow field entirely from scratch using the trained diffusion model, without imposing any specific conditions. For this purpose, we employed the gradient-based autoregressive method to produce long-span turbulence sequences spanning 300 time steps (equivalent to one flow-through time, as detailed in Table~\ref{table:idns_data}). This approach was chosen as the velocity tends to dampen in the replacement-based method during self-rollout, as previously mentioned. Figures~\ref{fig:idns:ucontour_uncondition}, \ref{fig:idns:vcontour_uncondition}, and \ref{fig:idns:wcontour_uncondition} display the generated instantaneous velocity fields alongside those simulated via DNS for comparison. Three randomly generated realizations of the spatiotemporal flow sequences are plotted, which all visually resample the DNS reference, showcasing our model's ability to produce diverse and realistic instantaneous velocity fields. This is fundamentally distinct from the approach using sequence neural networks (e.g., ConvLSTM or Transformer), which yields a single deterministic trajectory with a given initial condition. To further evaluate the generative performance, we compared the statistics of eight generated flow samples, each with 300 steps, against DNS data as shown in Fig.~\ref{fig:idns:mean_and_fluctuation_uncondition}. Remarkably, the mean streamwise velocity profile of the generated data mirrors that of the DNS across various flow regions, including the linear viscous sublayer, buffer layer, and logarithmic region. Furthermore, the root-mean-square (RMS) profiles of velocity fluctuations generated by our model demonstrate an high agreement with those obtained from DNS. This level of accuracy in capturing the essential turbulent characteristics underscores the robustness and versatility of our diffusion-based approach in simulating complex turbulent flows.
More statistics, including the spanwise energy spectra and two-point correlations of the velocity components at different wall distances, are provided in Appendix (see Figs.~\ref{fig:idns:energy_cas} and \ref{fig:idns:spatial_coorelation}). Overall, they are all reasonably in good agreement with those obtained from the DNS, especially in logarithmic region (e.g. $y^+=180$). Although slight difference in low-wave-number regions can be observed when probing the viscous sublayer and buffer layer, the anisotropic features along the wall direction are well captured.  Moreover, the two-point correlations of alll the generated flow samples agree with the DNS reference reasonable well, falling off to almost zero within a half width of the computational domain for both the streamwise and spanwise directions~\cite{abe2001direct}. There are small but noticeable bumps in the streamwise and spanwise two-point correction of our generated results in the near-wall regions, which is possibly due to large-scale isotropic structures generated around the channel center, which can be removed by introducing additional information using conditional sampling. 

\subsubsection{Conditioned generation given prior sequence}
\begin{figure}[tp]
	\centering
	\includegraphics[width=0.85\textwidth]{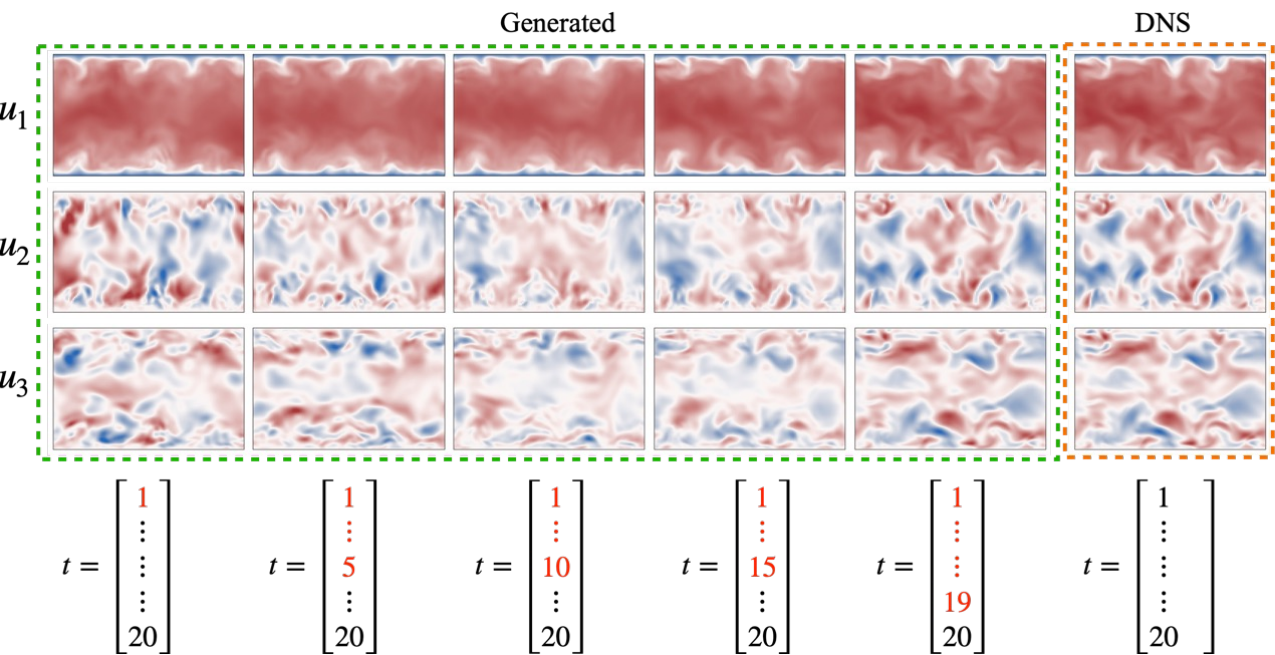}
	\caption{ The contours of streamwise ($u_1$) , wall-normal ($u_2$), spanwise ($u_3$) velocity at $t=20$ generated by diffusion model and DNS. In the context of $t$, the presence of red signifies conditioning, while black represents generation.  Each row use a same colorbar to highlight features in the corresponding velocity component. }
	\label{fig:idns:contour_hot_restart}
\end{figure}

In this section, we demonstrate the model's proficiency in generating conditioned sequences, a process where the model generates a sequence of flow based on a given set of prior flow sequence of varying lengths. To demonstrate this, we utilize an unseen sequence (spanning from $t = 1 \to t = 20$) from the DNS database. The length of the prior sequence that the diffusion model is conditioned on can be varied from 1 to 19. Namely, the diffusion model generates the remaining portion of the sequence up to $t = 20$ by sampling the trained model. Notably, the generation step is amplified a hundred times relative to the numerical step, resulting in a total sequence length of 2000 numerical steps. Fig.\ref{fig:idns:contour_hot_restart} presents the final snapshots of these generated sequences, compared against the last snapshot from the DNS simulations. The results show an expected trend: fewer prior snapshots lead to greater randomness in the final generated snapshot, while a higher number of prior snapshots results in a closer resemblance to the DNS counterpart. This trend of decreasing error, as illustrated in Fig.\ref{fig:idns:err_hot_restart}, effectively demonstrates the model's adeptness in incorporating conditional information and updating the posterior distribution accordingly. Moreover, it is crucial to highlight that these conditioned generations do not necessitate any retraining of the model. The diffusion model, once trained in an unconditional manner, can be directly employed for sampling with conditioning.

\newcommand{\pinkline}{\raisebox{2.0pt}{\tikz{\draw[-,pink,solid,line width = 1.5pt](0.,4.0mm) -- (5.5mm,4.0mm)}}}
\newcommand{\blackdash}{\raisebox{2.0pt}{\tikz{\draw[-,black,dashed,line width = 1.5pt](0.,4.0mm) -- (5.5mm,4.0mm)}}}

\subsubsection{Conditioned generation with desired statistics}
Our model has remarkable flexibility in conditioned generation, capable of producing flow with desired flow features. To demonstrate this capability, we consider generating instantaneous flow sequence given specified mean flow profiles or Reynolds stress fields. Unlike the previous conditional generation methods that rely on adding extra loss terms during the training phase~\cite{yousif2022physics}, our framework allows for incorporating these additional criteria directly into the inference (sampling) phase, eliminating the need for retraining.  This approach is feasible because the specified statistics are differentiable relative to the generated states. Such an approach not only saves the effort of repeated training but also adeptly addresses potential discrepancies between training and testing data, thereby validating the use of online conditioning over offline training. 

\begin{figure}[pt]
	\centering
	\includegraphics[width=0.9\textwidth]{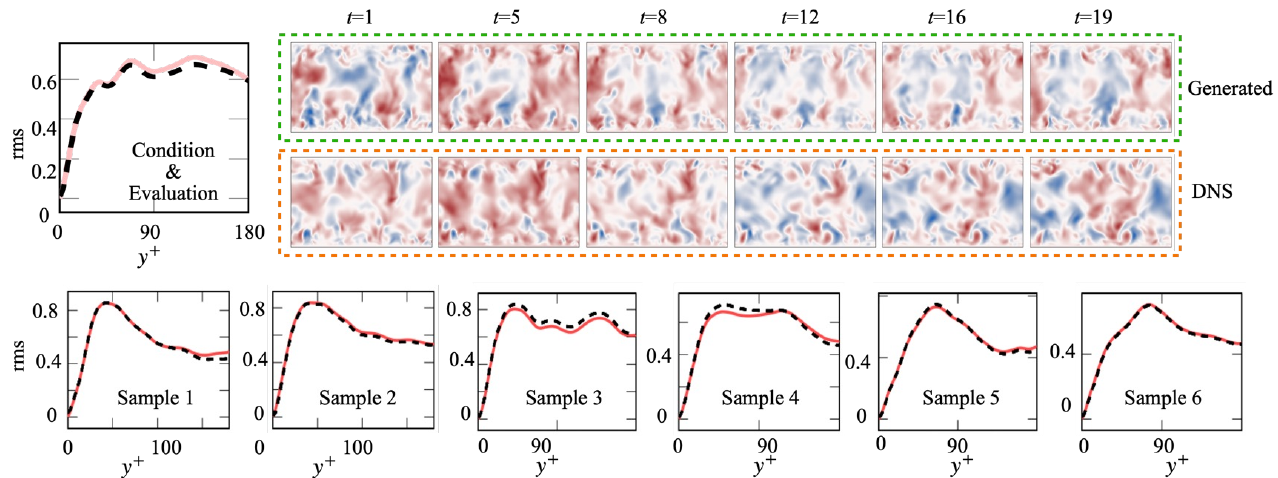}
	\caption{Conditional generation given specified RMS profiles of wall normal velocity fluctuation: The RMS profiles of wall-normal velocity fluction from the DNS (\protect\pinkline), and recalculated from the generated samples (\protect\blackdash).  Note that the sequences used to calculate RMS profiles are not in a statistically steady state, resulting in varying profiles across different sequences.}
	\label{fig:idns:vprime_guide}
\end{figure}

\underline{\textit{Conditioned on 1D mean flow profiles}}: 
As a proof of concept, we conditioned the model on the RMS profiles of wall-normal velocity fluctuation obtained from an unseen DNS sequence. As shown in Fig.~\ref{fig:idns:vprime_guide}, while our model generates distinct realizations, the RMS time-averaged fluctuation profile of the generated instantaneous flow agree well with that of the DNS data, suggesting that the condition is successfully imposed during the sampling process. Our model not only generates samples that visually similar to the DNS data but also can enforce the specified flow conditions without requiring retraining for any new mean flow profile requirements. We further tested our model by generating an additional six sequences using the same trained unconditional diffusion model, each conditioned on different RMS mean fluctuation profiles, and observed similarly robust performance (as shown in the lower part of Fig.\ref{fig:idns:vprime_guide}).

\begin{figure}[pt]
\centering
\includegraphics[width=0.9\textwidth]{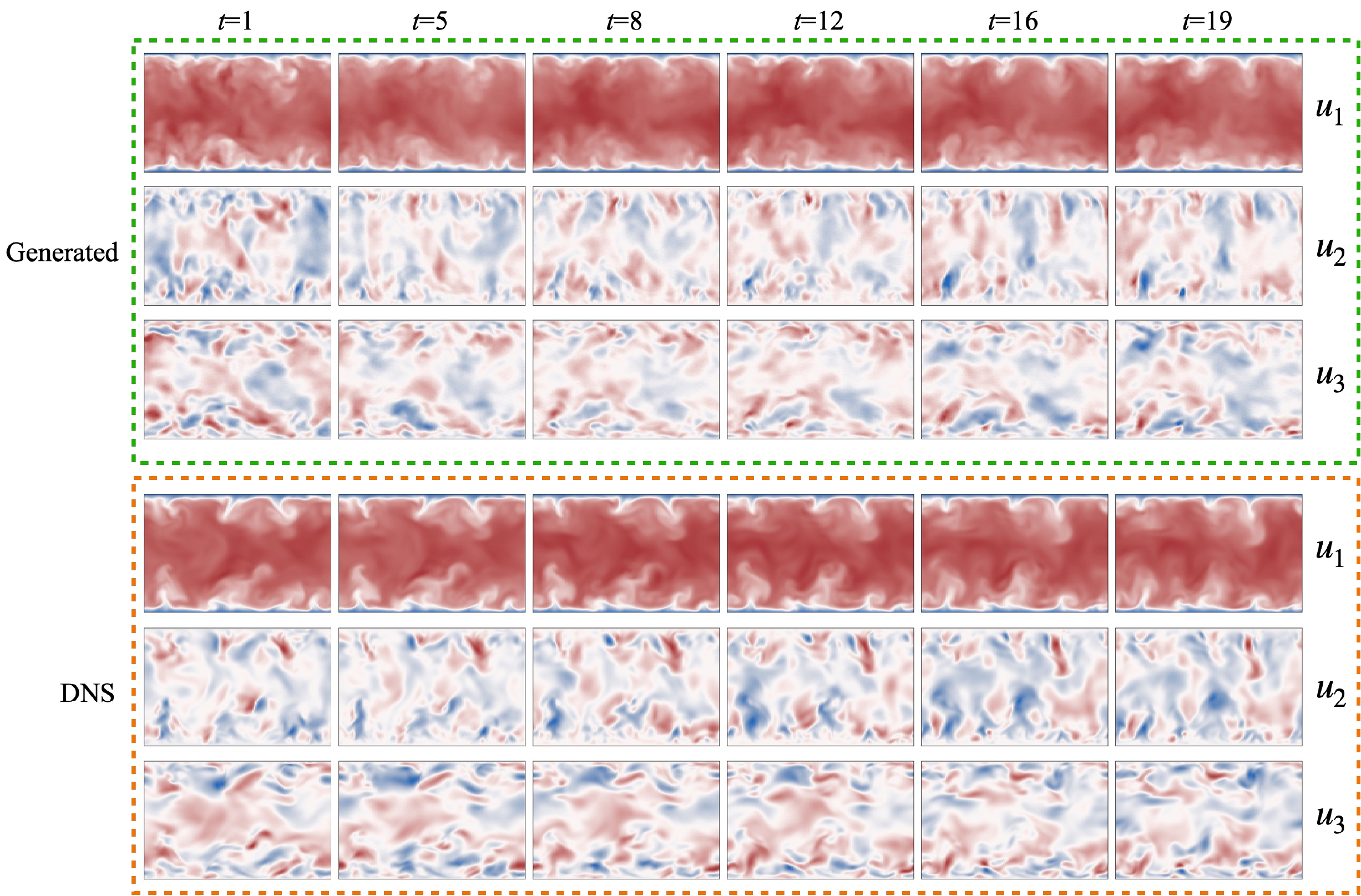}
%
\caption{The contours of streamwise ($u_1$), wall-normal ($u_2$), spanwise ($u_3$) velocity at the first, fifth, ninth, thirteenth, seventeenth, twentieth steps generated by diffusion model and DNS. Two models use a same colorbar for the same scalar field to highlight similarity in Reynolds stress and difference in flow fields.}
\label{fig:idns:Rss_guide_contour0}
\end{figure}

\begin{figure}[pt]
	\centering	
	\includegraphics[width=0.9\textwidth]{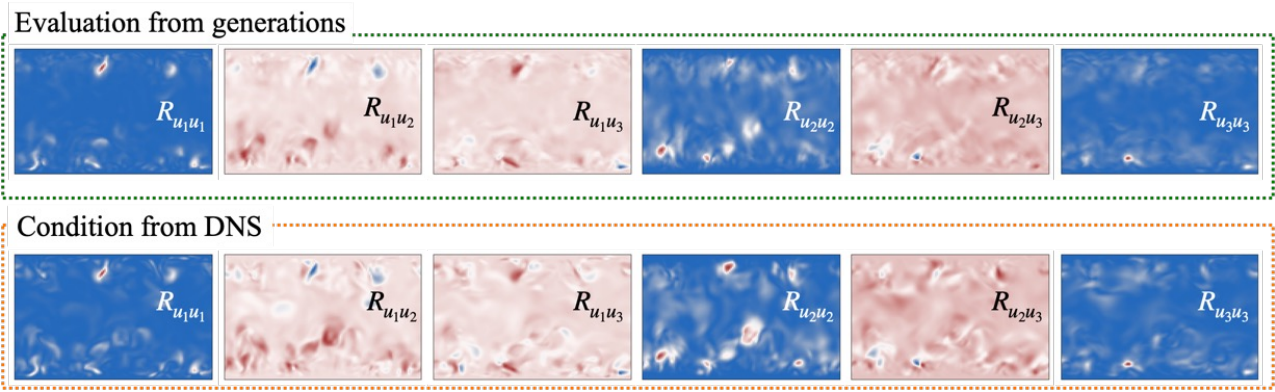}
	\caption{Re-evaluated and conditioned Reynolds stress between diffusion model  and DNS. Two models use a same colorbar for the same scalar field to highlight similarity in Reynolds stress and difference in flow fields.}
	\label{fig:idns:Rss_guide_contour1}
\end{figure}

\underline{\textit{Conditioned on the 2D Reynolds stress field}}: 
We next explored generating flow sequences with specified Reynolds stress fields using the same trained model. In contrast to the 1D RMS mean flow profiles, the Reynolds stress field contains more flow information but also requires a higher-dimensional representation. For this task, we first obtained the Reynolds stress from an unseen DNS flow sequence and then utilized the gradient-based conditional sampling method to generate a sequence of instantaneous flows that could lead to the same Reynolds stress field. As shown in Figs.~\ref{fig:idns:Rss_guide_contour0} and \ref{fig:idns:Rss_guide_contour1}, although the generated instantaneous velocity fields are significantly distinct from those in the DNS data, the six Reynolds stress tensor components computed from the generated flow are almost identical to those in the DNS data. This demonstrates the model's remarkable ability to synthesize instantaneous flow, while simultaneously ensuring the desired complex flow characteristics, such as Reynolds stresses. This capability underscores the potential of our model in simulating intricate turbulent dynamics, where adherence to certain statistical properties is crucial, yet a degree of unpredictability in the flow structures is maintained.

Finally, we briefly discuss the computational cost of different methods, emphasizing the efficiency of our approach. The diffusion model, serving as a surrogate sampler, distinct from traditional numerical simulations by directly generating instantaneous flow within a Bayesian sampling framework. This method not only achieves a remarkably good statistical match and facilitates flexible conditioning but also offers significant speed advantages over conventional simulation methods. As shown in Fig.~\ref{fig:idns:cost_compare}, in this channel flow case, the diffusion model running on a single GPU demonstrates a substantial speedup, being approximately 490 times faster than DNS executed using OpenFOAM~\cite{jasak2007openfoam} on CPUs. Even when compared to DNS conducted using our latest JAX-based, fully vectorized, GPU-enabled CFD solver~\cite{fan2023differentiable} on the same GPU, the diffusion model maintains a notable speedup, being around 16 times faster.

\subsection{Super-resolved generation of supersonic turbulent boundary layer}
\label{sec:duan}
The diffusion model, as a probabilistic modeling technique, offers flexibility in updating the distribution of high-resolution (HR) data based on low-resolution (LR) inputs to yield HR information. Traditional super-resolution (SR) techniques typically learn a deterministic mapping from a fixed input resolution. However, these techniques are not robust or would fail when the resolution of the inputs in inference phase is different from those in the training phase. Moreover, they struggle with extremely LR inputs, as the mapping may become ill-conditioned.  leading to ambiguities with multiple possible outputs for a single input. In this section, we explore the efficacy of the proposed diffusion model in generating flow fields from LR data for compressible, supersonic turbulent boundary layer (TBL) flow. Our SR process aims to generate the HR spatiotemporal flow field $\mathbf{u}(x, y, t) = [u(x, y, t), v(x, y, t), w(x, y, t), T(x, y, t)]^T: \Omega \times \mathbb{R}_{\geq0} \rightarrow \mathbb{R}^2$, conditioned on LR flow data, where $u, v, w, T$ are velocity components and temperature. Specifically, our focus is on examining the enhancement of flow patterns, statistical characteristics, and energy spectra with respect to the different resolution levels of the input LR data.

\subsubsection{Data preparation and model training}
We extract the 2D data (streamwise wall-normal plane) from a DNS of turbulent turbulent boundary layer flows at hypersonic speeds (Mach number $Ma = 6$) over a flat plate~\cite{zhang2018direct, duan2016pressure}. The computation domain spans $L_x \times L_y \times L_z=58.7\delta \times 15.7 \delta \times 39.7\delta$, where $\delta$ represents the boundary layer thickness ($\delta=13.8mm$). The original DNS resolution in streamwise cross-section is $1600 \times 500$.  For this study, we collect data from 25 distinct, non-overlapping streamwise sections, with each trajectory containing 160 time steps. To facilitate more efficient processing while retaining critical flow features, we downsampled the spatial resolution to $256\times256$ pixels. Our training dataset comprised 2400 snapshots, covering a broad range of flow feature. The model training was conducted once, and various SR tasks can be performed by using the gradient-based sampling conditioned on various LR inputs (refer to Section \ref{sec:gradient_method}). Note that we did not focus on generating long sequences in this experiment, as this capability has already been demonstrated in Section~\ref{sec:channel}. 


\subsubsection{SR generated fields given different low-resolution inputs}

\begin{figure}[pt]
	\centering	
	\includegraphics[width=1.0\textwidth]{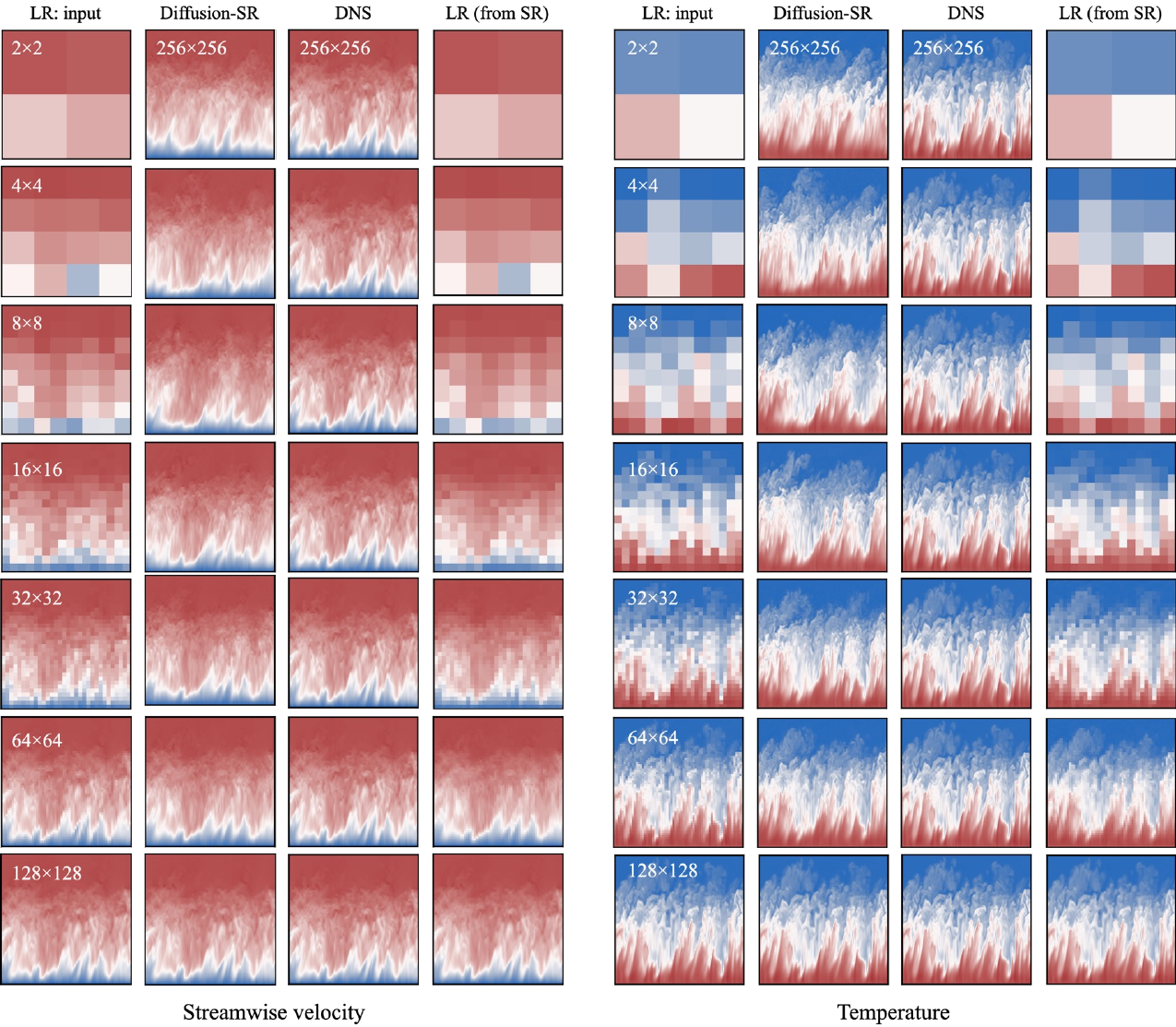}
	\caption{Instantaneous flow fields (streamwise velocity and temperature) of LR input, SR results, DNS and LR downsampled from SR results. The resolution of the LR input varies from $2\times2$, $4\times4$, $8\times8$, $16\times16$, $32\times32$, $64\times64$ and $128\times128$. The resolution of both SR results and DNS reference is $256\times256$. }
	\label{fig:duan:utcontour}
\end{figure}

Figures~\ref{fig:duan:utcontour} and~\ref{fig:duan:vwcontour} present the SR generated fields of streamwise, spanwise, wall-normal velocities, and temperature, conditioned on the LR inputs at different resolutions. From the visual comparison, it is apparent that the SR diffusion model is capable of enhancing the quality of the input data to produce HR snapshots that resemble the DNS reference to a remarkable degree, even when starting from LR inputs as coarse as $2 \times 2$. This suggests that our model possesses an advanced capability to generate the fine details necessary for an HR representation of the flow fields, which is particularly impressive at the lower resolutions. As we progress from the top row (lowest resolution of $2 \times 2$) to the bottom row (highest resolution of $128 \times 128$), there is a notable trend in the SR fields. At the lowest resolutions, the generated HR flow details of the SR output appear more randomized and are different from the DNS reference. As the resolution of the LR input increases, the SR-generated fields exhibit more of the fine structures and variations that are similar to those of the HR DNS reference. The last row ($128 \times 128$) showcases SR output that closely resembles the DNS reference, indicating that our SR diffusion model can generate that particular realization of the DNS instantaneous flow when it is conditioned on higher-resolution input data.
\begin{figure}[pt]
\centering
\input{pycode/duan/mean_var_uncondition_duan_2.tikz}
\caption{Fluctuation of streamwise, span-wise, wall-normal velocity and temperature, mean profile of streamwise velocity and temperature, energy spectrum (inside boundary layer $\frac{h}{H}=\frac{4}{256}$) of streamwise, span-wise, wall-normal velocity (\textit{from left to right, top to bottom}) from DNS  (256$\times$256, \ref{line:duan:rms_dns}), Diffusion-SR (256$\times$256, \ref{line:duan:rms_diff}) and Trilinear-SR (2$\times$2, \ref{line:duan:rms_dns_c}).}
\label{fig:duan:statistics:2by2}
\end{figure}
\begin{figure}[pt]
	\centering
	\input{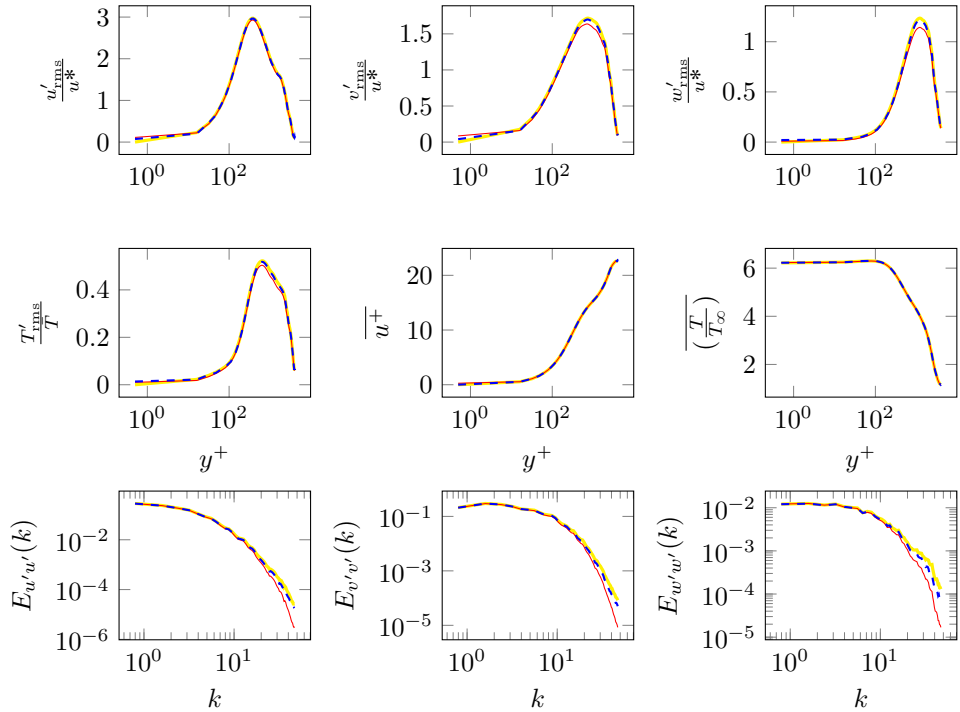}
	\caption{Fluctuation of streamwise, span-wise, wall-normal velocity and temperature, mean profile of streamwise velocity and temperature, energy spectrum (inside boundary layer $\frac{h}{H}=\frac{4}{256}$) of streamwise, span-wise, wall-normal velocity (\textit{from left to right, top to bottom}) from DNS  (256$\times$256, \ref{line:duan:rms_dns}), Diffusion-SR output (256$\times$256, \ref{line:duan:rms_diff}) and Trilinear-SR (128$\times$128, \ref{line:duan:rms_dns_c}).}
	\label{fig:duan:statistics:128by128}
\end{figure}
Moreover, we have conducted a comprehensive analysis of flow statistics and energy spectra, as depicted in Figs.~\ref{fig:duan:statistics:2by2} and~\ref{fig:duan:statistics:128by128} (results for other resolutions of LR inputs are shown in Figs.~\ref{fig:duan:statistics:4by4}, \ref{fig:duan:statistics:8by8}, \ref{fig:duan:statistics:16by16}, \ref{fig:duan:statistics:32by32}, and \ref{fig:duan:statistics:64by64}). Remarkably, even from very low-resolution inputs, the model successfully reconstructs the mean velocity and temperature profiles, which were initially indiscernible. The super-resolution process significantly enhances the fluctuating velocity and temperature profiles, especially in the near-wall regions. A critical metric of evaluation is the energy spectrum, which reflects the resolved turbulence scales. For comparative analysis, we upscaled the LR inputs to HR using trilinear interpolation before computing the energy spectrum. Although inputs at a very fine resolution of $128\times128$, the energy spectrum of trilinear-SR results agree with DNS results well, the discrepancy at high wave numbers are still notable. The proposed model adeptly bridges these gaps, facilitating precise and accurate recovery of the missing energy and turbulence scales. The energy spectra of the diffusion-SR results are almost identical to those of DNS reference, regardless of the input resolutions. This analysis reveals that our model effectively recovers this spectrum from LR inputs, indicating its efficiency in reproducing small-scale turbulence. It is important to note that our diffusion model is trained given DNS data once and then can be used for SR generation given different LR input resolutions without retraining. This is in stark contrast to previous SR works, which are highly reliant on the resolution used in training. Our model is much more robust and generalizable to LR input with different resolutions and qualities.

\section{Conclusion}
\label{sec:conclusion}

This study has introduced a novel Bayesian conditional diffusion model for generating spatiotemporal turbulent flows. Our proposed model unifies unconditional and conditional sampling strategies , offering a versatile solution across a spectrum of scenarios. We have systematically presented conditional sampling approaches that employs a gradient-based method for scenarios with a directly differentiable relationship between conditions and outcomes, and a replacement-based method for cases without such explicit relationship. Another novel aspect of our framework is its capability in effectively generating arbitrarily long sequences of turbulence flow through autoregressive gradient-based conditional sampling, negating the need for iterative retraining. Our empirical investigations underscore the model's adeptness in tackling a variety of turbulence generation tasks. The model exhibits prowess in synthesizing LES-like eddy-resolved turbulence from URANS inputs, producing turbulent channel flow sequences conditioned on desired flow statistics, and performing super-resolution generation on supersonic turbulent boundary layer flows from low-resolution input data with different resolutions and qualities. These capabilities not only demonstrate the model's versatility but also its potential to revolutionize the field of turbulence modeling and simulation.

Looking forward, we identify several promising trajectories for advancing this research. In scenarios involving differentiable conditions, enhancing estimation accuracy through refined sampling methodologies, such as importance sampling and particle filtering, remains a key area of interest. For non-differentiable conditions, exploring innovative strategies like reinforcement learning and ensemble Kalman filtering could yield significant advancements. Moreover, adapting our model to accommodate large-scale, graph-based data opens up new possibilities for tackling the complexities associated with general unstructured mesh data. Pursuing these avenues promises to yield substantial contributions to the domain of computational fluid dynamics and beyond.

\section*{Acknowledgments}
The authors would like to acknowledge the funds from Office of Naval Research under award numbers N00014-23-1-2071 and National Science Foundation under award numbers OAC-2047127. XF would also like to acknowledge the fellowship provided by the Environmental Change Initiative and Center for Sustainable Energy at University of Notre Dame. The content of this publication does not necessarily reflect the position or policy of any of these supporters, and no official endorsement should be inferred.

\appendix
\section{Supplementary results of flow over backward-facing step}
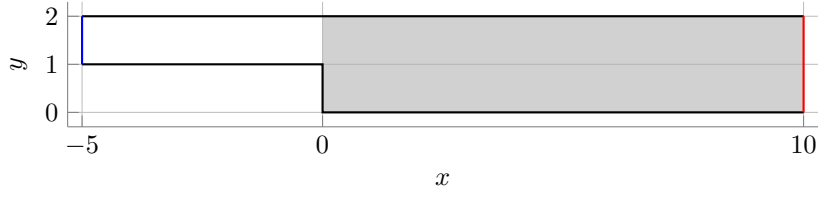
\begin{figure}[H]
	\centering
	\begin{tikzpicture}
\begin{axis}[
axis equal image,
axis line style={gray},
axis x line*=bottom,
axis y line*=left,
width=0.7\textwidth,
xtick={-5, 0, 10},
ytick={0, 1, 2},
grid=major,
ymax=2.3,
xmax=10.3,
ylabel=$y$,
xmin=-5.3,
xlabel=$x$,
ymin=-0.3]
\addplot [opacity=0.6, fill=black!30!white, opacity=0.6, forget plot]
coordinates {
( 0.00000000e+00,  0.00000000e+00)
( 1.00000000e+01,  0.00000000e+00)
( 1.00000000e+01,  2.00000000e+00)
( 0.00000000e+00,  2.00000000e+00)};

\addplot [thick, color=black]
coordinates {
(-5.00000000e+00,  1.00000000e+00)
( 0.00000000e+00,  1.00000000e+00)
( 0.00000000e+00,  0.00000000e+00)
( 1.00000000e+01,  0.00000000e+00)};\label{line:cyl0:wall}

\addplot [thick, color=black, forget plot]
coordinates {
(-5.00000000e+00,  2.00000000e+00)
( 1.00000000e+01,  2.00000000e+00)};

\addplot [thick, color=red]
coordinates {
( 1.00000000e+01,  0.00000000e+00)
( 1.00000000e+01,  2.00000000e+00)};\label{line:cyl0:out}

\addplot [thick, color=blue]
coordinates {
(-5.00000000e+00,  1.00000000e+00)
(-5.00000000e+00,  2.00000000e+00)};\label{line:cyl0:inlet}

\end{axis}
\end{tikzpicture}
	\caption{Geometry, domain of interest, and boundary conditions for the case of backward-facing step. Boundary conditions: inlet (\ref{line:cyl0:inlet}), outlet (\ref{line:cyl0:out}) and no-slip wall (\ref{line:cyl0:wall}). Shadow region is the domain of interest for generative modeling.}
	\label{fig:pitz_geo}
\end{figure}




\begin{figure}[H]
	\centering
	\begin{tikzpicture}
\begin{axis}[
width=0.5\textwidth,
xtick=\empty,
ymin=-40,
ylabel=wall-clock time (sec)]
\addplot [very thick, green, dashed]
coordinates {
( 8.50000000e-01,  0.00000000e+00)
( 8.50000000e-01,  5.90000000e+01)
( 1.15000000e+00,  5.90000000e+01)
( 1.15000000e+00,  0.00000000e+00)};\label{pitz:line:URANS_methd}

\node[below]    at    (axis cs:1.0, 0) {URANS};
\addplot [very thick, blue, solid]
coordinates {
( 1.85000000e+00,  0.00000000e+00)
( 1.85000000e+00,  3.10000000e+01)
( 2.15000000e+00,  3.10000000e+01)
( 2.15000000e+00,  0.00000000e+00)};\label{pitz:line:diff_methd}

\node[below]    at    (axis cs:2.0, 0) {vanilla};
\addplot [very thick, blue, solid]
coordinates {
( 2.85000000e+00,  0.00000000e+00)
( 2.85000000e+00,  1.73000000e+02)
( 3.15000000e+00,  1.73000000e+02)
( 3.15000000e+00,  0.00000000e+00)};\label{pitz:line:diff_methd}

\node[below]    at    (axis cs:3.0, 0) {gradient};
\addplot [very thick, blue, solid]
coordinates {
( 3.85000000e+00,  0.00000000e+00)
( 3.85000000e+00,  2.76000000e+02)
( 4.15000000e+00,  2.76000000e+02)
( 4.15000000e+00,  0.00000000e+00)};\label{pitz:line:diff_methd}

\node[below]    at    (axis cs:4.0, 0) {replace};
\addplot [very thick, red, solid]
coordinates {
( 4.85000000e+00,  0.00000000e+00)
( 4.85000000e+00,  3.77000000e+02)
( 5.15000000e+00,  3.77000000e+02)
( 5.15000000e+00,  0.00000000e+00)};\label{pitz:line:LES_methd}

\node[below]    at    (axis cs:5.0, 0) {LES};
\end{axis}
\end{tikzpicture}
	\caption{The wall-clock time for sampling a sequence of 240 time steps using variants of diffusion method (\ref{pitz:line:diff_methd}) and LES. The URANS (not as a sample method) here is provided for the reference.}
	\label{fig:pitz:cost_compare}
\end{figure}
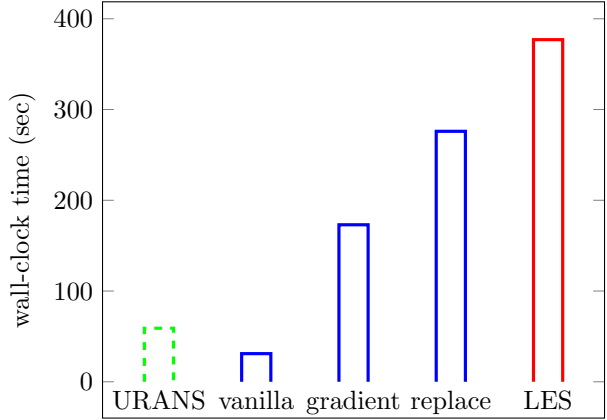

\begin{table}[H]
	\centering
	\begin{tabular}{|c | c | c | c| c| c| c|} 
		\hline
		& {\small Trubulence modeling} & Height & Width & \# step  & Inlet perturbation & Solver\\ 
		URANS & kEpsilon \cite{launder1983numerical} & 64 & 256& 240 & No & OpenFOAM \cite{jasak2007openfoam}\\ 
		LES   & dynamicK \cite{kim1995new}  & 64 & 256 & 240 & Yes  & OpenFOAM \cite{jasak2007openfoam}\\
		\hline
	\end{tabular}
	\begin{tabular}{|c | c | c | c| c| c| c| c|} 
		\hline
		Diff& $N_\mathrm{noise}$ & $N_\mathrm{inpaint}$ & $\sigma^\mathrm{guide}$ & $\beta^\mathrm{guide}$ & $N_\mathrm{previous}$ & $N_\mathrm{length}$ & device\\
		Vanilla& $20$ & None & None & None & $0$ & 40 & NVIDIA \\	
		Gradient& $20$ & $N_\mathrm{inpaint}$ & $(100,100)$ & $(0.02,0.02)$ & $20$ & $40$ & NVIDIA \\
		Replace& $20$ & 5 & None & None & $20$ & $40$ & NVIDIA \\
		\hline
	\end{tabular}
	\caption{The OpenFOAM is performed on AMD-Ryzen-9-7950, and diffusion model is evaluated on NVIDIA-GeForce-RTX-4090. In this case, we reduce the dimension of hyper-parameters by setting $\sigma^\mathrm{guide}_{N_\mathrm{noise}}=\dots=\sigma^\mathrm{guide}_{2}=\sigma^\mathrm{guide}(1)$, $\sigma^\mathrm{guide}_{1}=\sigma^\mathrm{guide}(2)$, $\beta^\mathrm{guide}_{N_\mathrm{noise}}=\dots=\beta^\mathrm{guide}_{2}=\beta^\mathrm{guide}(1)$, $\beta^\mathrm{guide}_{1}=\beta^\mathrm{guide}(2)$.}
	\label{table:les_rans_data}
\end{table}

\begin{figure}[H] 
	\centering
	\includegraphics[width=0.9\textwidth]{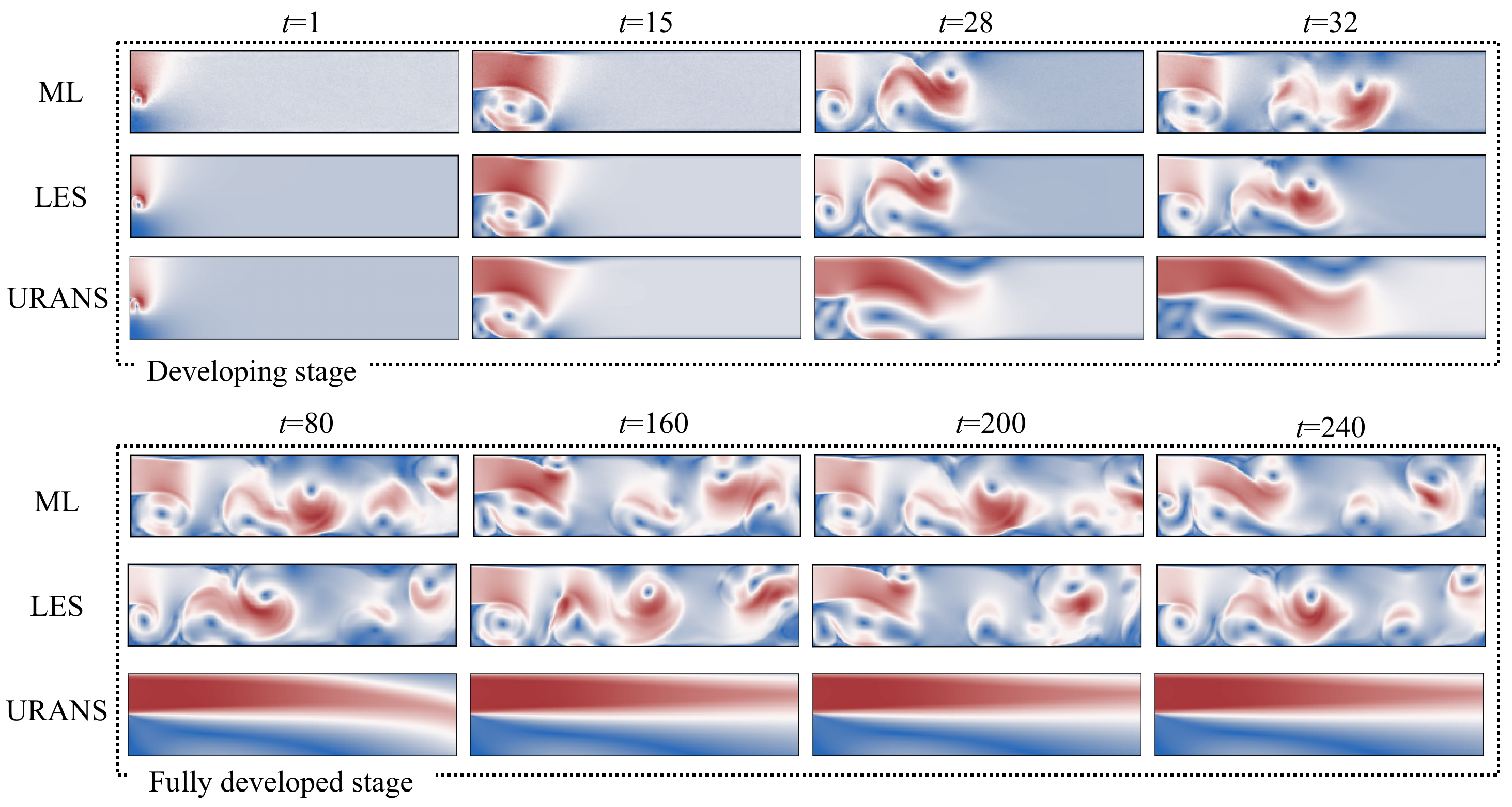}
	\caption{Velocity magnitude of generated samples (\textit{from first to fifth row}) and URANS (\textit{last row}) at $Re=13100$ from developing to fully-developed stage (\textit{from left to right})}
	\label{fig:pitz:vmag_Re13100}
\end{figure}

\begin{figure}[H]
	\centering
		\includegraphics[width=1.0\textwidth]{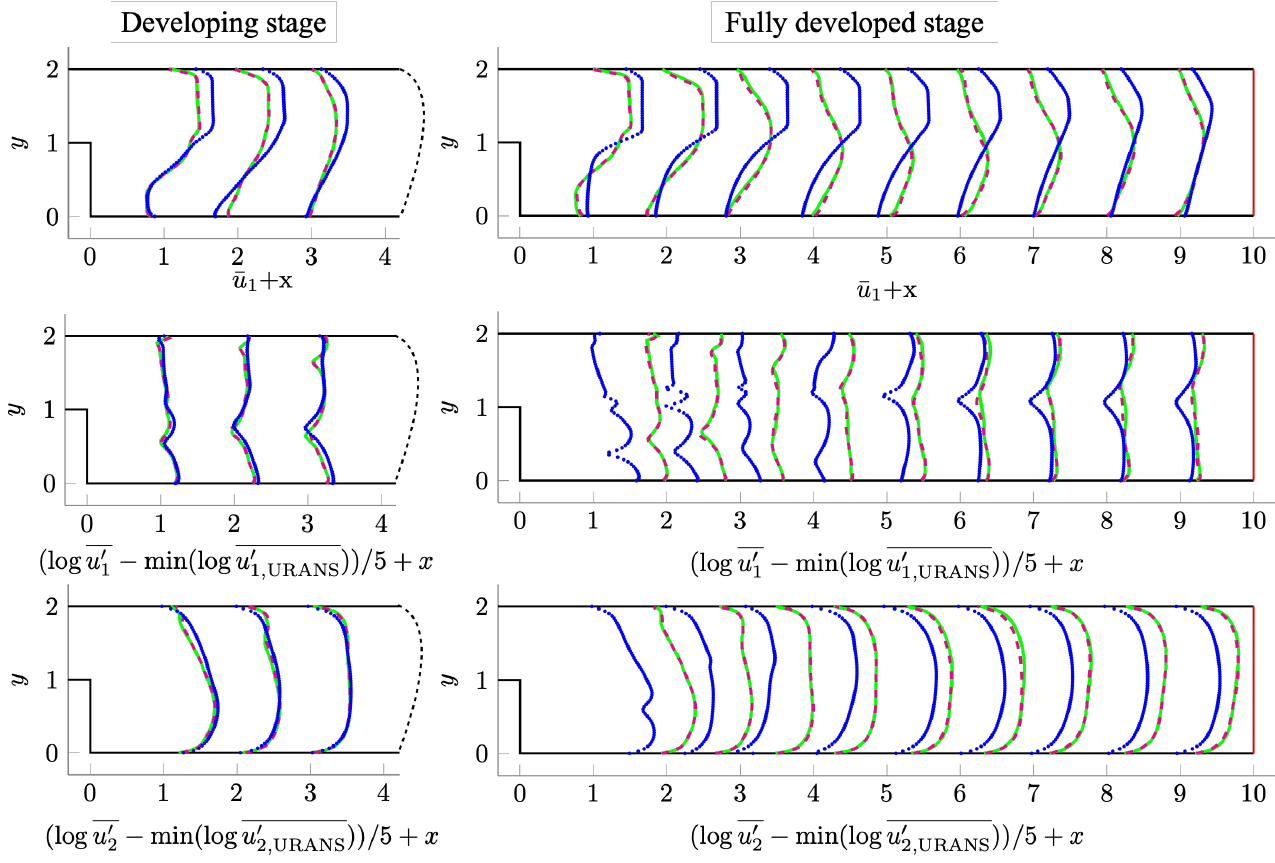}
	\caption{Mean of streamwise velocity, variance of streamwise velocity and variance of wall-normal velocity at $Re = 13100$ from LES (\protect\green), URANS (\protect\blue) and diffusion model (\protect\magenta).}
	\label{fig:pitz:umean_Re13100}
\end{figure}



\begin{figure}[H]
	\centering
	\input{pitz/compare_3_differnet_sample_with_noguide_var0.tikz}
	\caption{Variance of streamwise velocity at $Re=13100$ in the fully-developed phase: LES (\ref{line:pitz:mean_les}), URANS (\ref{line:pitz:mean_rans}), Diff-Gradient (\ref{line:pitz:mean_diff}), Diff-Replace (\ref{line:pitz:mean_replace}) and Diff-Vanilla (\ref{line:pitz:mean_vanilla}). The figures, arranged from left to right and top to bottom, correspond to locations at $x=[1,9]$.
	}
	\label{fig:pitz:uvar_Re13100_developed_compare_three_diff_method}
\end{figure}

\begin{figure}[H]
	\centering
	\input{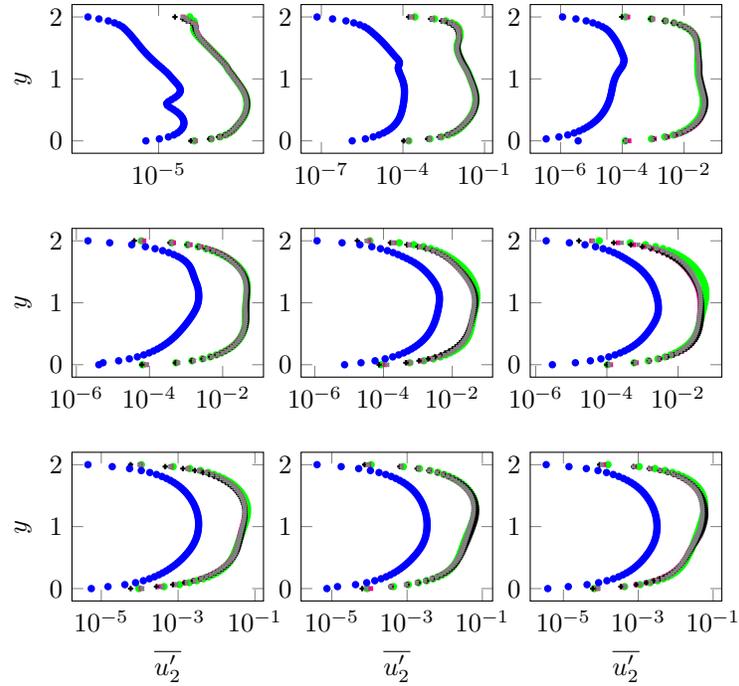}
	\caption{Variance of wall-normal velocity at $Re=13100$ in the fully-developed phase: LES (\ref{line:pitz:mean_les}), URANS (\ref{line:pitz:mean_rans}), Diff-Gradient (\ref{line:pitz:mean_diff}), Diff-Replace (\ref{line:pitz:mean_replace}) and Diff-Vanilla (\ref{line:pitz:mean_vanilla}). The figures, arranged from left to right and top to bottom, correspond to locations at $x=[1,9]$.
	}
	\label{fig:pitz:vvar_Re13100_developed_compare_three_diff_method}
\end{figure}

\section{Supplementary results of 3D turbulent channel flow}

\begin{figure}[H]
	\centering
	\input{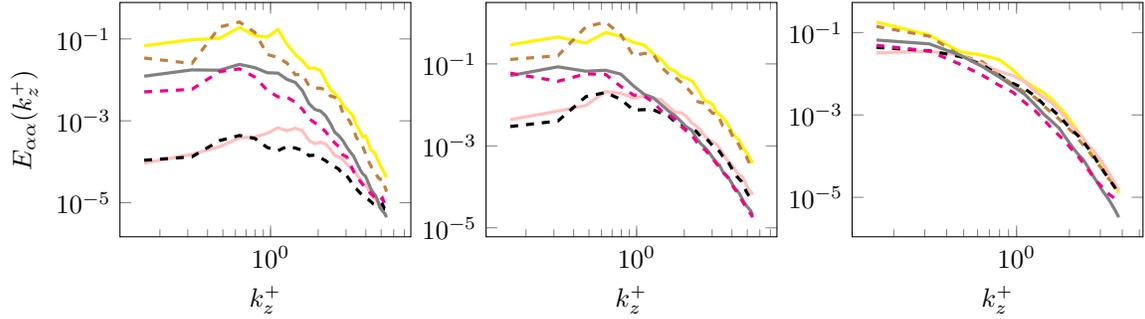}
	\caption{Energy spectra of the streamwise velocity from DNS (\ref{line:idns:uE_dns}), diffusion model (\ref{line:idns:uE_diff}), of the wall-normal velocity from DNS (\ref{line:idns:vE_dns}), diffusion model (\ref{line:idns:vE_diff}), of the spanwise velocity from DNS (\ref{line:idns:wE_dns}) and diffusion model (\ref{line:idns:wE_diff}) at $y^+=5$ (\textit{left}), $y^+=20$ (\textit{middle}) and $y^+=180$ (\textit{right}).
	}
	\label{fig:idns:energy_cas}
\end{figure}

\begin{figure}[H]
	\centering
	\input{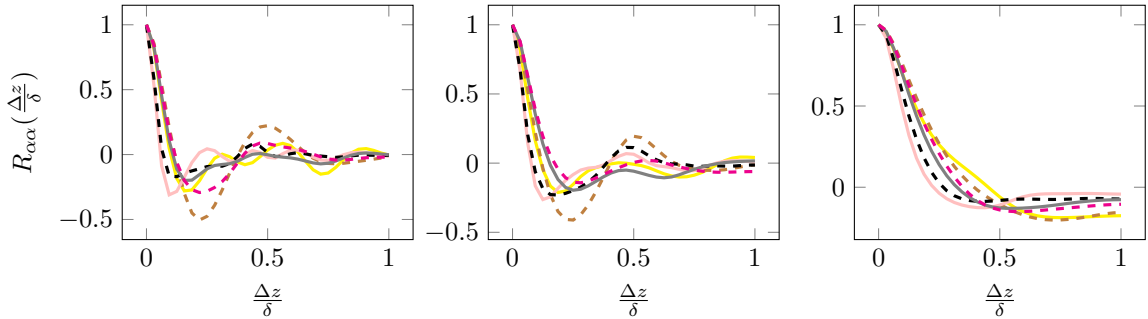}
	\caption{Spatial-correlations of the streamwise velocity from DNS (\ref{line:idns:uC_dns}), diffusion model (\ref{line:idns:uC_diff}), of the wall-normal velocity from DNS (\ref{line:idns:vC_dns}), diffusion model (\ref{line:idns:vC_diff}), of the spanwise velocity from DNS (\ref{line:idns:wC_dns}) and diffusion model (\ref{line:idns:wC_diff}) at $y^+=5$ (\textit{left}), $y^+=20$ (\textit{middle}) and $y^+=180$ (\textit{right}).
	}
	\label{fig:idns:spatial_coorelation}
\end{figure}

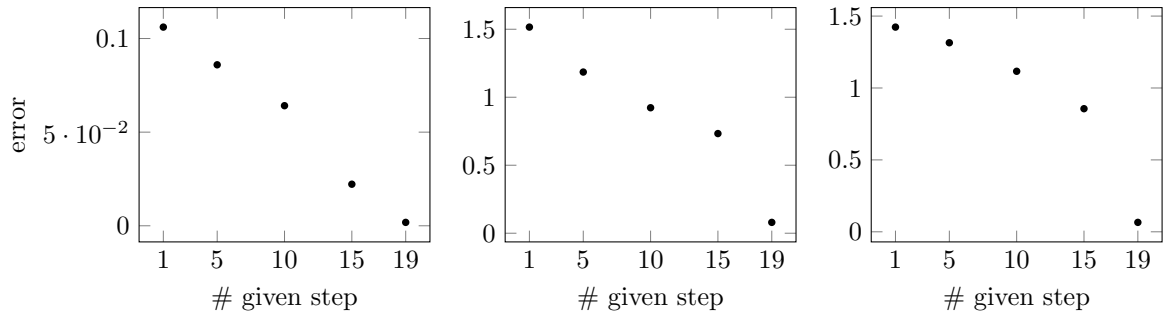
\begin{figure}[H]
	\centering
	\begin{tikzpicture}
\begin{groupplot} [
group style={group size = 3 by 1, horizontal sep = 1.0cm}]
\nextgroupplot[width=0.33\textwidth, xtick={1,5,10,15,19}, xlabel=\# given step, ylabel=error]
\addplot [mark options={solid, thick}, mark=*, mark size=1, only marks, forget plot]
coordinates {
( 1.00000000e+00,  1.06090471e-01)};

\addplot [mark options={solid, thick}, mark=*, mark size=1, only marks, forget plot]
coordinates {
( 5.00000000e+00,  8.59654620e-02)};

\addplot [mark options={solid, thick}, mark=*, mark size=1, only marks, forget plot]
coordinates {
( 1.00000000e+01,  6.41242415e-02)};

\addplot [mark options={solid, thick}, mark=*, mark size=1, only marks, forget plot]
coordinates {
( 1.50000000e+01,  2.21977569e-02)};

\addplot [mark options={solid, thick}, mark=*, mark size=1, only marks, forget plot]
coordinates {
( 1.90000000e+01,  1.81708334e-03)};

\nextgroupplot[width=0.33\textwidth, xtick={1,5,10,15,19}, xlabel=\# given step]
\addplot [mark options={solid, thick}, mark=*, mark size=1, only marks, forget plot]
coordinates {
( 1.00000000e+00,  1.51575494e+00)};

\addplot [mark options={solid, thick}, mark=*, mark size=1, only marks, forget plot]
coordinates {
( 5.00000000e+00,  1.18521845e+00)};

\addplot [mark options={solid, thick}, mark=*, mark size=1, only marks, forget plot]
coordinates {
( 1.00000000e+01,  9.22737598e-01)};

\addplot [mark options={solid, thick}, mark=*, mark size=1, only marks, forget plot]
coordinates {
( 1.50000000e+01,  7.32816219e-01)};

\addplot [mark options={solid, thick}, mark=*, mark size=1, only marks, forget plot]
coordinates {
( 1.90000000e+01,  7.97518268e-02)};

\nextgroupplot[width=0.33\textwidth, xtick={1,5,10,15,19}, xlabel=\# given step]
\addplot [mark options={solid, thick}, mark=*, mark size=1, only marks, forget plot]
coordinates {
( 1.00000000e+00,  1.42336190e+00)};

\addplot [mark options={solid, thick}, mark=*, mark size=1, only marks, forget plot]
coordinates {
( 5.00000000e+00,  1.31457114e+00)};

\addplot [mark options={solid, thick}, mark=*, mark size=1, only marks, forget plot]
coordinates {
( 1.00000000e+01,  1.11618102e+00)};

\addplot [mark options={solid, thick}, mark=*, mark size=1, only marks, forget plot]
coordinates {
( 1.50000000e+01,  8.56008232e-01)};

\addplot [mark options={solid, thick}, mark=*, mark size=1, only marks, forget plot]
coordinates {
( 1.90000000e+01,  6.55208603e-02)};

\end{groupplot}\end{tikzpicture}
	\caption{The error of streamwise (\textit{left}), wall-normal (\textit{middle}), spanwise (\textit{right}) velocity at the twenty step generated by diffusion model given the first, the first five, the first ten, the first fifteen and the first nineteen steps.}
	\label{fig:idns:err_hot_restart}
\end{figure}

\begin{table}[H]
	\centering
	\begin{tabular}{| c  | c| c| c| c| } 
		\hline
		 Solver & Height & Width   \\ 
		 {OpenFOAM} \cite{jasak2007openfoam}     & 256 & 128   \\ 
		 {FanCFD}   \cite{fan2023differentiable} & 256 & 128   \\
		\hline
	\end{tabular}
	
	\begin{tabular}{|c | c | c | c| c| c| c|} 	
		\hline
		Sampling& $N_\mathrm{noise}$ & $N_\mathrm{inpaint}$ & $\sigma^\mathrm{guide}$ & $\beta^\mathrm{guide}$ & $N_\mathrm{previous}$ & $N_\mathrm{length}$ \\
		Unconditional & $20$ & None & (100,100) & (0.018,0.018) & $10$ & 20  \\
		Condition on initial condition & $20$ & 5 & None & None & $1-19$ & 20 	\\
		{Condition on rms } & $20$ & None & (100,100) & (0.015,0.015) & None & 20 	\\
		{Condition on Reynolds stress} & $20$ & None & (100,100) & (0.013,0.013) & None & 20 	\\
		\hline
		
	\end{tabular}
	\caption{In this case, we reduce the dimension of hyper-parameters by setting $\sigma^\mathrm{guide}_{N_\mathrm{noise}}=\dots=\sigma^\mathrm{guide}_{2}=\sigma^\mathrm{guide}(1)$, $\sigma^\mathrm{guide}_{1}=\sigma^\mathrm{guide}(2)$, $\beta^\mathrm{guide}_{N_\mathrm{noise}}=\dots=\beta^\mathrm{guide}_{2}=\beta^\mathrm{guide}(1)$, $\beta^\mathrm{guide}_{1}=\beta^\mathrm{guide}(2)$.}
	\label{table:idns_data}
\end{table}

\begin{figure}[H]

	\includegraphics[width=0.95\textwidth]{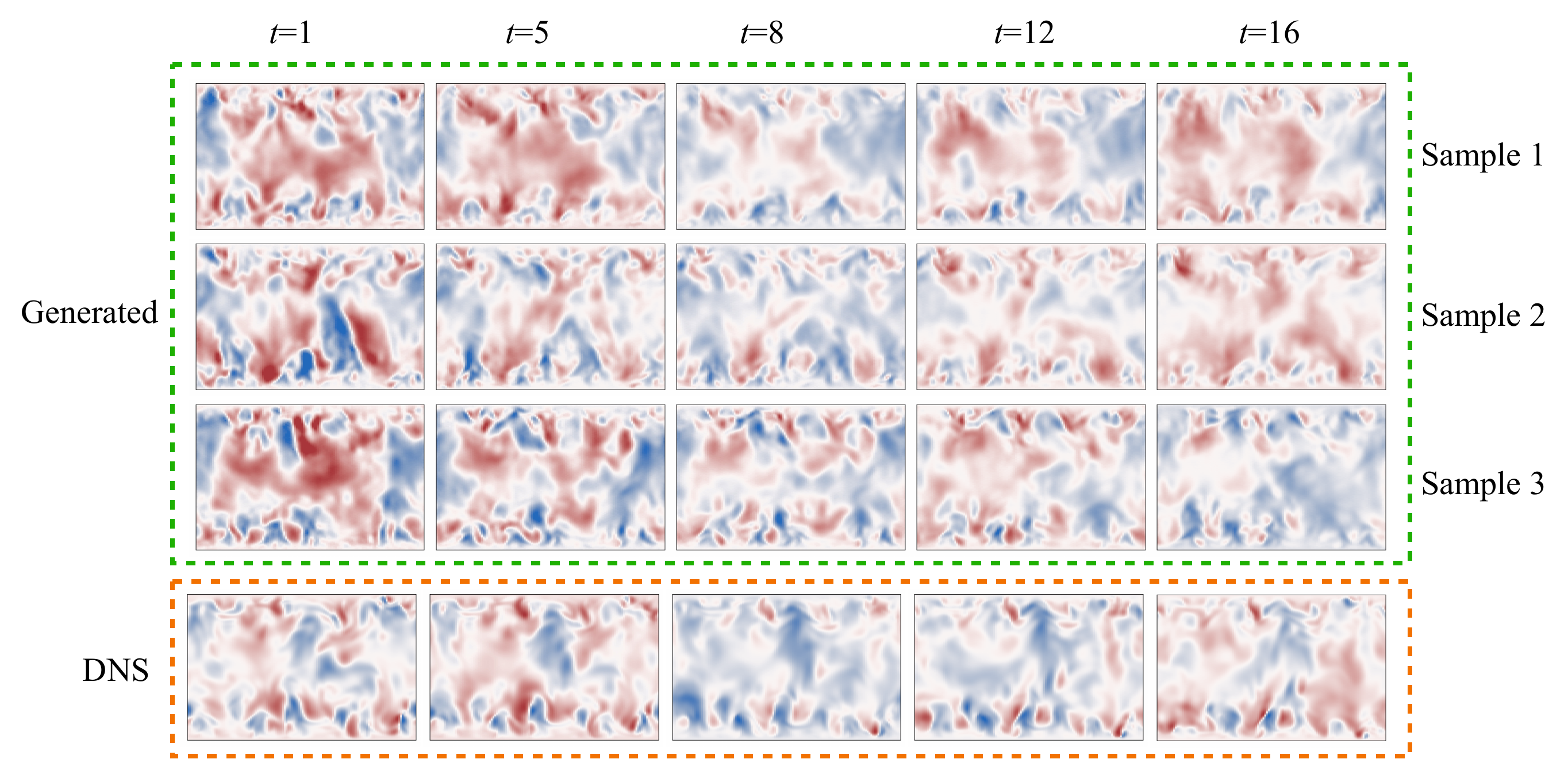}

	\caption{Continuous instantaneous wall-normal velocity fields of samples from diffusion model and DNS.}
	\label{fig:idns:vcontour_uncondition}
\end{figure}

\begin{figure}[H]

	\includegraphics[width=0.95\textwidth]{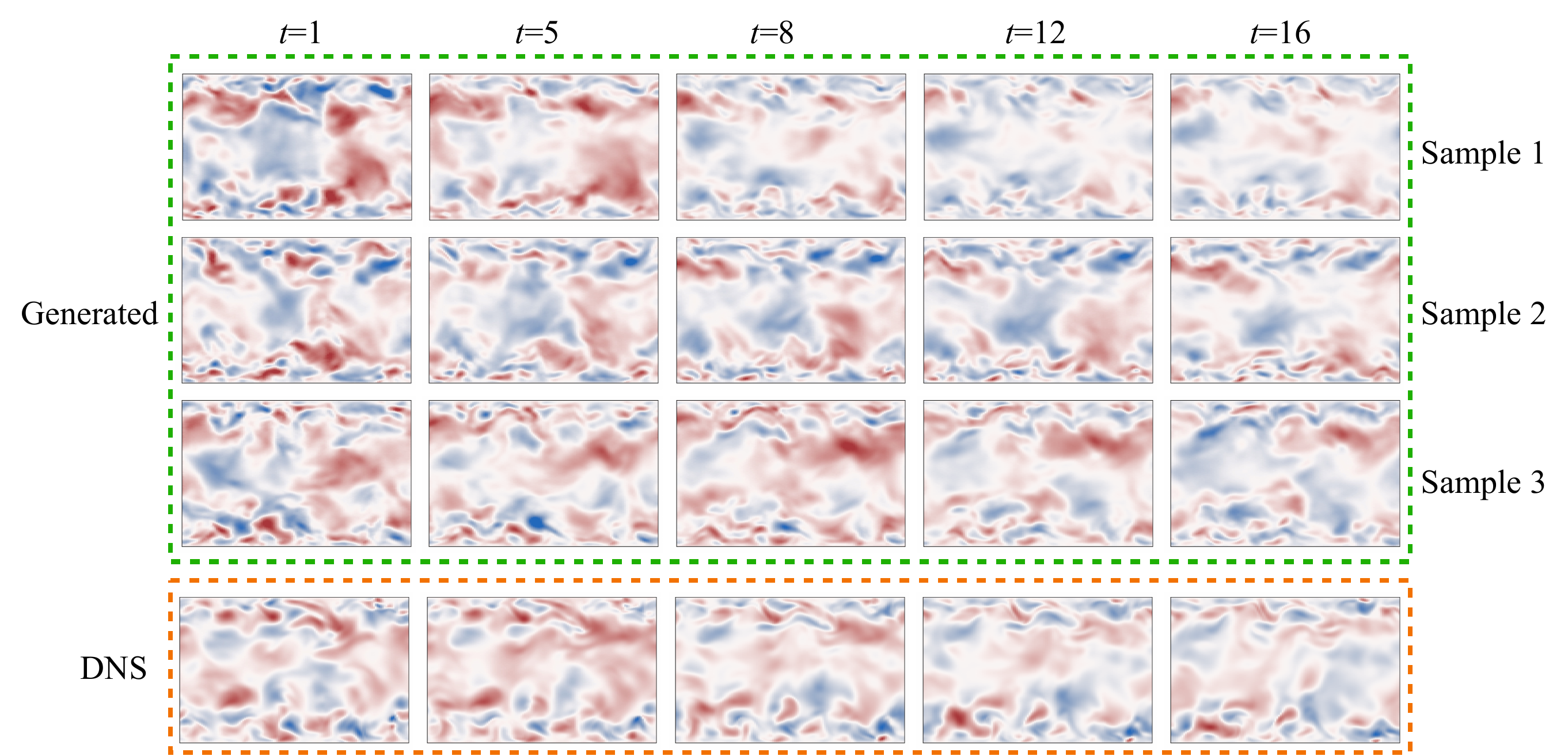}

	\caption{Continuous instantaneous spanwise velocity fields of samples from diffusion model and DNS.}
	\label{fig:idns:wcontour_uncondition}
\end{figure}

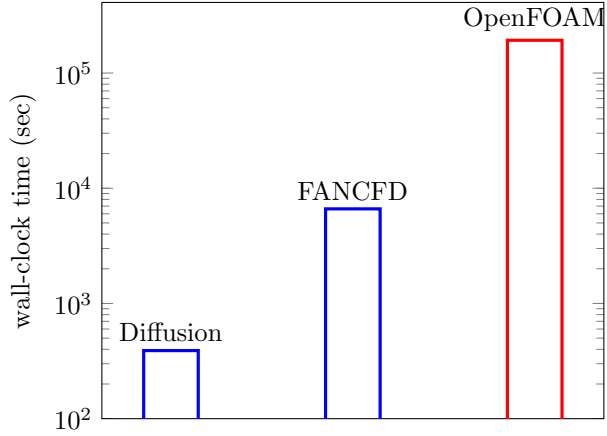
\begin{figure}[H]
	\centering
	\begin{tikzpicture}
\begin{axis}[
width=0.5\textwidth,
xtick=\empty,
ytick={1,10,100,1000,10000,100000},
ylabel=wall-clock time (sec),
ymode=log,
ymin=100]
\addplot [very thick, blue, solid]
coordinates {
( 8.50000000e-01,  1.00000000e+01)
( 8.50000000e-01,  3.90000000e+02)
( 1.15000000e+00,  3.90000000e+02)
( 1.15000000e+00,  1.00000000e+01)};\label{idns:line:diff_methd}

\node[above]    at    (axis cs:1.0, 390) {Diffusion};
\addplot [very thick, blue, solid]
coordinates {
( 1.85000000e+00,  1.00000000e+01)
( 1.85000000e+00,  6.62400000e+03)
( 2.15000000e+00,  6.62400000e+03)
( 2.15000000e+00,  1.00000000e+01)};\label{idns:line:gpu}

\node[above]    at    (axis cs:2.0, 6624) {FANCFD};
\addplot [very thick, red, solid]
coordinates {
( 2.85000000e+00,  1.00000000e+01)
( 2.85000000e+00,  1.92492000e+05)
( 3.15000000e+00,  1.92492000e+05)
( 3.15000000e+00,  1.00000000e+01)};\label{idns:line:cpu}

\node[above]    at    (axis cs:3.0, 192492) {OpenFOAM};
\end{axis}
\end{tikzpicture}
	\caption{The wall-clock time for sampling a flowthrough (a sequence of 300 time steps) using the gradient-diffusion method and DNS methods. Specifically, OpenFOAM \cite{jasak2007openfoam} is performed on CPU (\ref{idns:line:cpu}, Intel Xeon(R) Gold 6138) with 16 cores in parallel; FANCFD \cite{fan2023differentiable} and diffusion are performed on single GPU card (\ref{idns:line:gpu}, NVIDIA-GeForce-RTX-4090).}
	\label{fig:idns:cost_compare}
\end{figure}

\section{Supplementary results of supersonic turbulent boundary layers}

%
%
%
%
%
%

\begin{table}[H]
	\centering
	\begin{tabular}{| c  | c| c|} 
		\hline
		Solver & Height & Width   \\ 
		DuanCFD \cite{zhang2018direct}     & 256 & 256 \\ 
		\hline
	\end{tabular}
	
	\begin{tabular}{|c | c | c | c| c| c| c|} 	
		\hline
		Input resolution& $N_\mathrm{noise}$ & $N_\mathrm{inpaint}$ & $\sigma^\mathrm{guide}$ & $\beta^\mathrm{guide}$ & $N_\mathrm{refine}$ & $N_\mathrm{length}$ \\
		2$\times$2     & $10$ & None & (100,100) & (0.015,0.001) & 10  & 10 \\
		4$\times$4     & $10$ & None & (100,100) & (0.015,0.001) & 10  & 10 	\\
		8$\times$8     & $10$ & None & (100,100) & (0.01,0.001) & 10 & 10 	\\
		16$\times$16   & $10$ & None & (100,100) & (0.02,0.02) & 1 & 10 	\\
		32$\times$32   & $10$ & None & (100,100) & (0.05,0.02) & 1 & 10 	\\
		64$\times$64   & $10$ & None & (100,100) & (0.06,0.02) & 1 & 10 	\\
		128$\times$128 & $10$ & None & (100,100) & (0.08,0.0035) & 120 & 10 \\
		\hline
		
	\end{tabular}
	\caption{In this case, we reduce the dimension of hyper-parameters by setting $\sigma^\mathrm{guide}_{N_\mathrm{noise}}=\dots=\sigma^\mathrm{guide}_{2}=\sigma^\mathrm{guide}(1)$, $\sigma^\mathrm{guide}_{1}=\sigma^\mathrm{guide}(2)$, $\beta^\mathrm{guide}_{N_\mathrm{noise}}=\dots=\beta^\mathrm{guide}_{2}=\beta^\mathrm{guide}(1)$, $\beta^\mathrm{guide}_{1}=\beta^\mathrm{guide}(2)$.}
	\label{table:duan_data}
\end{table}

\begin{figure}[H]
	\includegraphics[width=1.0\textwidth]{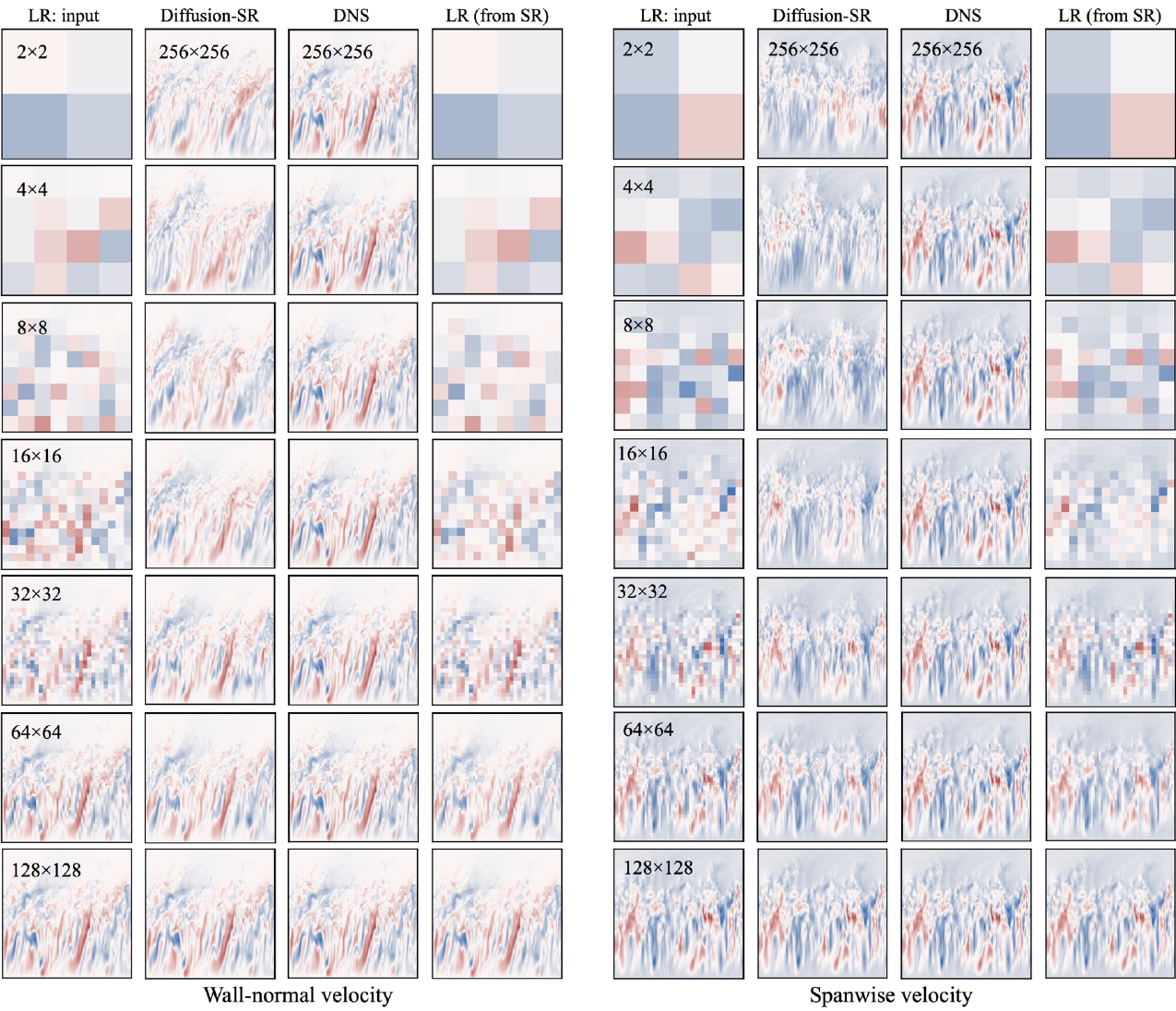} 
	\caption{Instantaneous flow fields (span-wise and wall-normal velocity) of LR input, SR results, DNS and LR downsampled from SR results. The resolution of the LR input varies from $2\times2$, $4\times4$, $8\times8$, $16\times16$, $32\times32$, $64\times64$ and $128\times128$. The resolution of both SR results and DNS reference is $256\times256$. }
	\label{fig:duan:vwcontour}
\end{figure}

\begin{figure}[H]
	\centering
	\input{pycode/duan/mean_var_uncondition_duan_4.tikz}
	\caption{Fluctuation of streamwise, span-wise, wall-normal velocity and temperature, mean profile of streamwise velocity and temperature, energy spectrum (inside boundary layer $\frac{h}{H}=\frac{4}{256}$) of streamwise, span-wise, wall-normal velocity (\textit{from left to right, top to bottom}) from DNS  (256$\times$256, \ref{line:duan:rms_dns}), SR result (256$\times$256, \ref{line:duan:rms_diff}) and LR input (4$\times$4, \ref{line:duan:rms_dns_c}).}
	\label{fig:duan:statistics:4by4}
\end{figure}

\begin{figure}[H]
	\centering
	\input{pycode/duan/mean_var_uncondition_duan_8.tikz}
	\caption{Fluctuation of streamwise, span-wise, wall-normal velocity and temperature, mean profile of streamwise velocity and temperature, energy spectrum (inside boundary layer $\frac{h}{H}=\frac{4}{256}$) of streamwise, span-wise, wall-normal velocity (\textit{from left to right, top to bottom}) from DNS  (256$\times$256, \ref{line:duan:rms_dns}), SR result (256$\times$256, \ref{line:duan:rms_diff}) and LR input (8$\times$8, \ref{line:duan:rms_dns_c}).}
	\label{fig:duan:statistics:8by8}
\end{figure}

\begin{figure}[H]
	\centering
	\input{pycode/duan/mean_var_uncondition_duan_16.tikz}
	\caption{Fluctuation of streamwise, span-wise, wall-normal velocity and temperature, mean profile of streamwise velocity and temperature, energy spectrum (inside boundary layer $\frac{h}{H}=\frac{4}{256}$) of streamwise, span-wise, wall-normal velocity (\textit{from left to right, top to bottom}) from DNS  (256$\times$256, \ref{line:duan:rms_dns}), SR result (256$\times$256, \ref{line:duan:rms_diff}) and LR input (16$\times$16, \ref{line:duan:rms_dns_c}).}
	\label{fig:duan:statistics:16by16}
\end{figure}

\begin{figure}[H]
	\centering
	\input{pycode/duan/mean_var_uncondition_duan_32.tikz}
	\caption{Fluctuation of streamwise, span-wise, wall-normal velocity and temperature, mean profile of streamwise velocity and temperature, energy spectrum (inside boundary layer $\frac{h}{H}=\frac{4}{256}$) of streamwise, span-wise, wall-normal velocity (\textit{from left to right, top to bottom}) from DNS  (256$\times$256, \ref{line:duan:rms_dns}), SR output (256$\times$256, \ref{line:duan:rms_diff}) and LR  input (32$\times$32, \ref{line:duan:rms_dns_c}).}
	\label{fig:duan:statistics:32by32}
\end{figure}

\begin{figure}[H]
	\centering
	\input{pycode/duan/mean_var_uncondition_duan_64.tikz}
	\caption{Fluctuation of streamwise, span-wise, wall-normal velocity and temperature, mean profile of streamwise velocity and temperature, energy spectrum (inside boundary layer $\frac{h}{H}=\frac{4}{256}$) of streamwise, span-wise, wall-normal velocity (\textit{from left to right, top to bottom}) from DNS  (256$\times$256, \ref{line:duan:rms_dns}), SR output (256$\times$256, \ref{line:duan:rms_diff}) and LR input (64$\times$64, \ref{line:duan:rms_dns_c}).}
	\label{fig:duan:statistics:64by64}
\end{figure}


\end{document}